\documentclass[letterpaper,12pt]{article} 
\usepackage{fontenc}
\usepackage{epsfig}
\usepackage{amssymb}
\usepackage{amsmath}
\usepackage{rotating}
\usepackage{setspace}
\usepackage{bm}
\epsfverbosetrue
\def\abstract{\if@twocolumn
\section*{Abstract}
\else \normalsize 
\begin{center}
{\bf Summary\vspace{-.5em}\vspace{0pt}} 
\end{center}
\quotation 
\fi}
\def\endabstract{\if@twocolumn\else\endquotation\fi}

\makeatletter
\@addtoreset{equation}{section}
\makeatother

\newtheorem{remark}{Remark}[section]

\newcommand{\myappendix}[1]{
	\setcounter{section}{1}
        \renewcommand{\thesection}{A\arabic{section}}}

\begin{document}
\pagestyle{empty}
\doublespacing
\begin{titlepage}
\title{Econometric Modeling of Regional
Electricity Spot Prices in the Australian Market}
\author{Michael Stanley Smith$^{1}$ and Thomas S. Shively$^{2,\star}$}
\date{This Version: April 2018}
\maketitle
{\small
\noindent
1. Melbourne Business School, University of Melbourne,
200 Leicester Street, Carlton, VIC 3053, Australia; Email: mike.smith@mbs.edu\\
2. Department of Information, Risk, and Operations Management,
McCombs School of Business, University of Texas at Austin,
1 University Station, B6500, Austin, TX 78712-0212, U.S.A; 
Email: Tom.Shively@mccombs.utexas.edu
}
\vspace{2in}

\noindent {\small
$^{\star}$ Corresponding Author}\\

\noindent {\small 
This
work was partially supported by
Australian Research Council Future Fellowship FT110100729.}

\newpage
\begin{center}
{\LARGE 
Econometric Modeling of Regional
Electricity Spot Prices in the Australian Market}\\
\vspace{5pt}
{\bf Abstract}
\end{center}
\noindent
\onehalfspacing
Wholesale electricity markets are increasingly integrated 
via high voltage interconnectors, and inter-regional
trade in electricity is growing. To model this, 
we consider a spatial
equilibrium model of price formation, where constraints on inter-regional
flows result in three distinct equilibria in prices. We use this to motivate
an econometric model for the distribution of observed electricity spot prices
that captures many of their unique empirical characteristics.
The econometric model features
supply and inter-regional
trade cost functions, which are estimated using Bayesian monotonic 
regression smoothing methodology.
A copula multivariate time series model is employed
to capture additional
dependence --- both cross-sectional and serial --- in regional
prices. The marginal distributions are nonparametric, with
means given by the regression means. The model has the advantage of
preserving
the heavy right-hand tail in the predictive densities of price.
We fit the model to half-hourly spot
price data in the
five interconnected regions of the
Australian national electricity market.
The fitted model is then used to measure how both supply and
price shocks in one region are transmitted to the distribution of
prices in all regions in subsequent periods.
Finally, to validate our econometric model, we show that
prices forecast using the proposed model compare favorably
with those from some benchmark alternatives.
\vspace{2.5in}



\end{titlepage}
\doublespacing

\newpage
\pagestyle{plain}
\newpage
\section{Introduction}
\vspace{-10pt}
During the past two decades, traditional
vertically integrated electricity power systems have been replaced with wholesale
markets in many countries and regions.
While the design of such markets varies, day-ahead and bid-based markets are common.
In such markets, bids are placed
by generators, distributing utilities and
third parties at an intraday resolution one day prior
to consumption (or `dispatch'). These auctions are overseen by network management organizations, which
schedule the dispatch of electricity on a least 
cost basis, subject to demand forecasts and system security requirements.
The spot price is the
highest priced bid
dispatched to meet demand at a point in time.
The 
effective modeling and forecasting of spot prices is important for utilities to operate in
a market profitably, and also for the management of a market; see Kirschen \& Strbac~(2004)
for an introduction to contemporary electricity markets.

In North America, Australia and New Zealand,
markets also feature nodal pricing, where there are different prices in different
regions within a market.
Electricity is traded
between regions via high voltage interconnectors, and increased
synchronization of regional prices is often considered symptomatic of increased market 
efficiency. Moreover, even when nodal pricing is not a market feature,
the construction of interconnectors
between adjacent power systems allows for trade between markets with different prices.
In this study, we propose a new approach for the modeling
of the joint distribution of regional
electricity spot prices in a bid-based interconnected market. A
statistical model is proposed that
is motivated by an economic spatial equilibrium pricing model. We aim 
to show
that exploiting the
structural relationships
between regional supply, interconnector flows and regional prices suggested
by the economic model
helps explain
some of the unique empirical features of electricity prices.
It also provides a framework
from which density forecasts can be constructed and event studies undertaken.

Electricity is a flow commodity that cannot be stored economically. Demand has a
strong predictable component --- for example, see 
Harvey \& Koopman~(1993), Ramanathan et al.~(1997),
Smith~(2000), Taylor et al.~(2006), Clements et al.~(2016) and many others
--- and is almost perfectly 
inelastic to the wholesale price in the short run because most consumers face fixed tariffs.
Supply comprises generators with very different marginal costs
 of production and 
`ramping' times (i.e. the 
time required to change the amount generated). Trade between regions is
also constrained by interconnector capacity. These features ensure that time series of
spot prices
exhibit unique characteristics, including sharp spikes,
 asymmetry, periodicity, and 
cross-sectional
and serial dependence; all of which have been documented widely
(Knittel \&~Roberts~2005; Panagiotelis \& Smith~2008;
Karakatsani \& Bunn~2008b; Clements et al.~2017).
In response to this complexity, a variety of nonlinear
statistical models have been developed for modeling and forecasting
intraday spot prices; see Weron~(2014) for an extensive
overview. Of these, regime switching regression and time series
models have been particularly successful (Huisman \& Mahieu~2003; Weron et al.~2004;
Haldrup \& Neilsen~2006; Karakatsani \& Bunn~2008b; Janczura \& Weron~2010; 
Bunn et al.~2016), where
different latent regimes correspond to different price distributions under
different economic equilibria. Yet less attention has been
given to the multivariate modeling of regional prices, which is what we 
undertake here.

Our approach is motivated by an
extension of the spatial equilibrium model of DeVany \& Walls~(1999). This
is a network economic model (Nagurney~1999) that relates each
regional supply curve (often called a `stack' in the power systems literature) and
inter-regional transmission cost functions, to prices in all regions.
It features
three inter-regional price equilibria.
The first is where prices are synchronized across regions,
up to transmission costs, 
the second is where prices deviate due to temporal ramping constraints, and the 
third is where prices deviate due to interconnector capacity constraints; the latter scenario
is often
called `congestion pricing' in the power systems literature (Bakirtzis~2001). 
However, this economic
model
cannot be implemented directly in practice. One reason is that it is written
in terms of directed electricity flows between all pairwise 
combinations of regions, which are
unobservable in wholesale pools. Another reason is that aggregate
cost and supply functions are difficult to construct for many markets.

We therefore consider a statistical approximation, where
pairwise flows are replaced with observed directed flows on inter-regional interconnectors,
and cost and supply functions are estimated using the Bayesian
monotonic function estimation 
methodology of Shively et al.~(2009). 
We extend the approach of these authors to allow
the regression disturbances to follow a mixture of three normals, with moments
corresponding to those of the
price distributions under the three theoretical equilibria.
To allow for additional 
serial and cross-sectional dependence in prices
we employ
a multivariate time series model based on a copula construction
of the type discussed
by Biller~(2009), Smith~(2015) and Smith \& Vahey~(2016). 
Such dependence is likely to arise due to
strategic bidding,
the impact of intermittent renewable generation, the effect of climatic conditions on the power system,
and other market imperfections.
The marginal mean of each price series is determined by the monotonic regression means, 
and the mean-corrected prices are modeled using their empirical distribution functions, so that 
each price series distribution is marginally nonparametric. 
Both serial and cross-sectional
dependence is then captured
by a high-dimensional Gaussian copula, with
parameter matrix equal
to the autocorrelation matrix of a
stationary vector autoregression (VAR) process. The VAR has a long lag 
length, but with many null coefficient matrices, so that it is still
parsimonious. The autocorrelation matrix of the VAR can be readily estimated
using sparse matrix computations, and 
provides an estimate of the Gaussian copula parameter matrix.

Copulas are popular tools for multivariate modeling because they 
allow dependence to be captured separately from the other features
of the distribution, which 
are in turn captured by arbitrary marginal models. They are used
increasingly 
in the energy modeling literature, with applications as diverse as accounting
for spatial dependence in renewable generation
(Papaefthymiou \& Kurowicka~2009), hedging of electricity futures
(Liu et al.~2010) and modeling energy spreads (Westgaard, 2014; Ch.4).
Copulas are also attractive for the
multivariate modeling of regional electricity prices,
because they allow for
extreme levels of asymmetry and heavy tails through an appropriate 
choice of marginal model. For example,
Smith et al.~(2012), Wang, Cai \& He~(2015), 
and Ignatieva \& Tr\"uck~(2016) all use various low-dimensional
copulas to capture cross-sectional dependence between electricity
prices in different regions, while Manner, T\"urk \& Eichler~(2016)  use low-dimensional
copulas to capture the cross-sectional dependence between price
spike incidence in different regions.
However, our use of copulas differs
from these authors in that we use a much higher dimensional
copula to capture both serial and cross-sectional dependence. 
Such a copula-based time series model
preserves the nonparametric
marginal distributions of prices, including the high levels of
asymmetry, heavy tails and regression function means.

We outline how to construct the joint predictive density of regional prices from the fitted model using forecasts of electricity load,
which are readily available in practice; for example, 
see Clements et al.~(2016) and references therein. 
The network economic model motivates a 
different multivariate time series model for each supply region, each
producing
a separate joint predictive distribution of prices in all regions. 
We combine these distributions to produce an ensemble forecast distribution.
Ensemble methods for combining predictive densities from different models have 
proven both popular and useful in the fields of
meteorology (Sloughter et al.~2010) and 
macroeconometrics (Mitchell \& Hall~2005; Jore et al.~2010).

We apply our model to the Australian National Electricity Market (NEM) using
half-hourly
data. This market is the world's longest
interconnected power system, with five regions that 
each have their own price setting mechanisms.
The market design
shares features with many other wholesale markets, and the price series exhibit
the empirical characteristics often observed in such markets. Estimates
of the supply functions are highly nonlinear, and in line with those
from
previous studies; for example, see Geman \& Roncoroni~(2006).
We show how the fitted models can be used to assess the impact
on prices in all regions
of a supply-side shock in just one region, 
such as the loss of a major generating unit. We also
compute the generalized impulse response (Koop, Pesaren \& Potter~1996)
of the nonlinear multivariate time series model to assess how 
regional price shocks are transmitted to the distribution of prices in all
regions in subsequent periods.
In addition, to validate our model we show that its predictions
are more accurate than two na\"ive benchmarks.

We note here that our approach has a number of 
features that are similar to those found in some
other recent statistical models of electricity prices.
These include quantile regression and time series
models (Bunn et al.~2016;
J\'onsson et al.~2014), which
are popular for this problem because they
produce flexible forecast densities.
By employing nonparametric marginal distributions,
this is also the case for the forecast densities produced by 
our copula multivariate time series model. However,
it is considerably more complex to extend 
quantile regression to the multiple time
series case we focus on here.
Regression
or time series models
that include `fundamental'
supply and/or demand-side variables
have also proven popular for forecasting electricity prices; see, for 
example, Karakatsani \& Bunn~(2008a; 2008b),
Ferkingstad et al.~(2011) and
Gonzalez et al.~(2012). Our
regression models with monotonic effects in supply and 
inter-regional flows also employ fundamental supply-side variables, 
motivated 
by the structural relationships in the network economic model.
Moreover, by exploiting an economic model, our study
is similar in spirit to that of Ziel and Steinert~(2016), who construct dynamic
demand and supply curves based on a time series model fit using
auction data.
Last,
constructing ensemble forecast densities from five separate
multivariate statistical time series models for regional prices, mirrors the long-standing
practice of point forecast combination used previously
in the energy literature; for example, see De Menezes et al.~(2000).

The rest of the paper is organized as follows. Section~\ref{sec:mod} outlines the network
economic model, and the statistical model for prices.
Section~\ref{sec:nem} introduces the NEM and the data. 
Section~\ref{sec:em} discusses estimation of the monotonic
regressions, 
and the copula model. The density forecasting methods are also
discussed here. Section~\ref{sec:emp} contains the empirical analysis, 
including the two event studies and
the validation study, while Section~\ref{sec:discuss} concludes.

\section{Modeling Electricity Prices}\label{sec:mod}

\subsection{Spatial Equilibrium Pricing}\label{sec:sep}
We follow DeVany \& Walls~(1999) and consider a spatial equilibrium
model of price formation. We denote the independent price-setting
regions in an inter-connected electricity market
as nodes $\Omega=\{1,2,\ldots,r\}$. Let $\Gamma = \Omega \times \Omega$ 
be the set of
origin-destination
(O/D) pairs, between which there is an energy flow of $Q_{i,j}\geq 0$ from
node $i$ to node $j$. In general, the flow $Q_{i,j}$ 
is not directly observable in an electricity network, because
it consists of the sum of flows down all different pathways from node $i$
to node $j$ in the power system. Nevertheless, 
the electricity demanded $d_j$ at node $j$, and the electricity 
supplied $b_i$ at node $i$, are both observed. 
Under the assumption of zero transmission loss, the
feasibility conditions are
\[
d_j = \sum_{i\in \Omega} Q_{i,j}\;\mbox{ and }\;
b_i = \sum_{j\in \Omega} Q_{i,j}\,,
\]
and
the total energy generated and consumed is equal with $\sum_{i\in \Omega}b_i=
\sum_{j\in \Omega}d_j$. 

In the absence of load-shedding, demand for electricity in most bid-based
markets
is almost perfectly inelastic with respect to price at any instant
in time
(Kirschen~2003). This is because wholesale prices are not passed
through directly to the consumer, who largely face fixed tariffs. We denote the 
monotonically
increasing inverse supply function at node $i$
as $S_i(b_i)$. 
In a perfect market, the following spatial equilibrium conditions 
(Nagurney~1999, Chap. 3.1) hold to clear
the market for all O/D pairs $(i,j)\in \Gamma$:
\begin{equation}
S_i(b_i)+C_{i,j}(Q_{i,j})
\left\{\begin{array}{ccl} = \pi_j &\mbox{if} &Q_{i,j}>0 \\ 
\geq \pi_j &\mbox{if} &Q_{i,j}=0 \end{array}\right. \,,
\label{eq:equil0}
\end{equation}
where $\pi_j$ is the demand price at node $j$, and $C_{i,j}(Q_{i,j})$ is
the cost of flow $Q_{i,j}$. The monotonically increasing cost
function $C_{i,j}$ 
measures the wheeling costs
due to transmission losses, congestion in the transmission system and possibly
aspects of
market design. In the above, we
refer to the origin node $i$ as the supply region.

Flows between nodes are constrained by capacity constraints on 
the interconnectors, so that there is also a constraint
$\bar Q_{i,j}$ on flow $Q_{i,j}$. In the presence of this constraint,
the spatial equilibrium conditions (Nagurney~1999, Chap 3.3) at 
Equation~(\ref{eq:equil0}) are now
\begin{equation}
S_i(b_i)+C_{i,j}(Q_{i,j})
\left\{\begin{array}{ccl} = \pi_j &\mbox{if} &0<Q_{i,j}<\bar Q_{i,j} \\ 
\geq \pi_j &\mbox{if} &Q_{i,j}=0 \\
\leq \pi_j &\mbox{if} &Q_{i,j}=\bar Q_{i,j} \end{array}\right. \,,
\label{eq:equil1}
\end{equation}
for all O/D pairs. The first of the two inequalities occurs when
there are no gains from trade because the supply price plus wheeling
costs at node $i$,  $S_i(b_i)+C_{i,j}(Q_{i,j})$, exceeds the price
$\pi_j$ received at
node $j$. For example, this
can occur in practice when baseline generators are supplying
electricity at node $j$ at a low price because there are
additional costs to ramping down
the generators; see Wolak~(2007) for a discussion of the impact of ramping costs on 
market clearing conditions. 
The second inequality occurs when trade is
constrained because of capacity constraints on inter-regional interconnectors. 
In practice, in many bid-based markets,
this can result in a price spike at node $j$,
because to meet demand
at this node electricity has to be supplied locally, and the inverse supply function
typically 
has a first derivative that is monotonically increasing.
For a depiction of a stylized inverse supply 
function, see
Geman \& Roncoroni~(2006).

\subsection{Statistical Model of Regional Prices}\label{sec:smrp}
In bid-based markets, electricity prices
are set at
equally-spaced points in time at an intra-day resolution. It is
difficult to employ
the spatial equilibrium at Equation~(\ref{eq:equil1}) to directly
model these prices
for a number of reasons.
First, as flows between nodes increase, the energisation of an electricity network 
is governed by Kirchoff's voltage law (Wood \& Wollenberg~1995; Krischen \& Strbac~2004). 
This determines how the currents (and therefore the 
power flows) distribute themselves through a network, in a manner that
is not necessarily efficient economically. 
Second, in a centralized bid-based
wholesale market, while flows are observable on individual interconnectors,
flows $Q_{i,j}$ between
individual node pairs cannot be distinguished.\footnote{Note that 
$Q_{i,j}$ can be observed in theory in markets with bilateral trades, although
these are rare in practice, and do not occur in the Australian NEM.}
Third, the upper bounds $\bar Q_{i,j}$ also
depend on any other flows that share transmission pathways.
Fourth, extension of the static model at Equation~(\ref{eq:equil1})
to a dynamic situation is difficult, in part due to
differing ramping times and costs for different generation capacity.

Therefore, we instead use the spatial equilibrium model in Section~\ref{sec:sep}
to motivate a statistical model for
electricity prices.
Each supply region produces a separate multivariate
model for prices observed at all price-setting
nodes in the network. We outline the model for a given supply region $i$ below,
and it has two main components.
The first is a series of regression models for
prices, each of which has
a finite mixture distribution for the disturbances. The second is
a copula
to capture
both cross-sectional and serial dependence in prices.

Let $E_{i,j}$ be a
set of directed arcs that connect the O/D pair $(i,j)$ that
corresponds to the physical interconnectors between
the two nodes. If there are no such interconnectors, then $E_{i,j}=\emptyset$.
Let
$v_{a}\geq 0$ be the flow on directed arc $a$, and 
$c_a$ be a monotonically increasing transmission cost
function on the bounded interval
$[0,\bar{v}_a]$, with $c_{a}(0)=0$. In our statistical model, the
cost function
for the unobservable flow $Q_{i,j}$ is replaced
by the sum of the costs of flows along the arcs in $E_{i,j}$, so that
$C_{i,j}(Q_{i,j})\approx \sum_{a\in E_{i,j}} c_a(v_a)$. 
At time $t$, let
$\pi_{j,t}$ be the price in region $j$, $b_{i,t}$ the supply
in region $i$, and $v_{a,t}$ the flow on directed arc $a$. Then,
the 
equilibrium conditions at Equation~(\ref{eq:equil1}) motivate
the following regression 
for observations at times $t=1,\ldots,T$:
\begin{eqnarray}
\pi_{j,t} &= &S_i(b_{i,t})+\sum_{a \in E_{i,j}}c_a(v_{a,t})+\epsilon_{i,j,t}
\label{eq:stat1}\\
 &= &\eta_{i,j,t}+\epsilon_{i,j,t}\nonumber\,.
\end{eqnarray}
There are separate regressions
for each
combination
$(i,j)\in \Gamma$, so that there are $r^2$ regressions in total. We
refer to the regression above as the region $i$ supply regression for
prices in region $j$.
The random variable $\epsilon_{i,j,t}$ is marginally distributed as
a mixture of
three distributions,
with distribution function
\begin{equation}
F^{\tiny \epsilon}_{i,j}(\epsilon)=\sum_{l=1}^3 \omega_{i,j,l} G_l(\epsilon;\alpha_{i,j,l},\sigma_{i,j,l})\,.
\label{eq:mixdist}
\end{equation}
Here, the
weights $0<\omega_{i,j,l}<1$ are such that 
$\omega_{i,j,1}+\omega_{i,j,2}+\omega_{i,j,3}=1$, and
$G_l(\cdot;\alpha,\sigma)$ is a distribution function with 
mean $\alpha$ and standard deviation $\sigma$.
The marginal mean $E(\epsilon_{i,j,t})=\sum_{l=1}^3 \omega_{i,j,l} \alpha_{i,j,l}
=\tilde \alpha_{i,j}$ is therefore
nonzero, and the marginal mean of prices is $E(\pi_{j,t})=\eta_{i,j,t}+\tilde \alpha_{i,j}$.

The adoption of this mixture distribution is motivated by 
the three equilibria in the spatial model.
It is also consistent with
previous empirical analyses
that detect two or
three regimes in time series of prices;
for example, see Karakasani \& Bunn~(2008b) and Janczura \& Weron~(2010).
To identify the first two moments of the three mixture components,
we assume the following parameter constraints:
\begin{itemize}
\item[] {\em Component 1}: $\alpha_{i,j,1}$ and $\sigma_{i,j,1}$ are unconstrained\quad ({\em baseline case});
\item[] {\em Component 2}: $\alpha_{i,j,2}<\alpha_{i,j,1}$ and $\sigma_{i,j,2}>\sigma_{i,j,1}$\quad ({\em lower mean and higher variance});
\item[] {\em Component 3}: $\alpha_{i,j,3}>\alpha_{i,j,1}$ and $\sigma_{i,j,3}>\sigma_{i,j,1}$\quad ({\em higher mean and higher variance}).
\end{itemize}
Component~1 corresponds to the case where $0<Q_{i,j}<\bar Q_{i,j}$
in Equation~(\ref{eq:equil1}). 
Component~2 corresponds to the case
where $Q_{i,j}=0$, which can occur when high costs to ramping down
large baseline generators produce lower (even negative)
prices $\pi_j$ with 
higher variability in comparison to Component~1.
Component~3 corresponds to the case where $Q_{i,j}=
\bar Q_{i,j}$, and prices $\pi_j$ are both higher and more
volatile in comparison to Component~1. 
This last component includes situations that lead to price spikes; 
something that is observed empirically in many bid-based electricity markets, 
including the Australian market. While these moment constraints 
are motivated by 
the features of the three equilibria, they also identify the likelihood so that
the problem of label-switching (e.g. see Fr\"uhwirth-Schnatter~2006; Sec. 3.5) 
does not occur in their estimation.
 
A multivariate model for
the impact
of supply in region $i$ on prices in all regions 
$\bm{\pi}_t=(\pi_{1,t},\ldots,\pi_{r,t})'$ over times $t=1,\ldots,T$
can be constructed using a copula function. Copulas are popular tools
to account for cross-sectional dependence in non-Gaussian models
(Patton~2006),
and also serial dependence in time series
(Smith et al.~2010, Smith~2015). They allow the adoption of 
arbitrary marginal models, followed by the modeling
of dependence in a second separate step.
Here, we employ the regression models above as the marginal
models for the $r$ elements of $\bm{\pi}_t$. 
We employ a Gaussian
copula (Song~2000) to capture both cross-sectional and serial
dependence in prices over-and-above that explained by supply and interconnector
flows, as discussed below.

For each supply region $i$, we use a copula representation of the
distribution of all observed 
prices $\bm \pi=(\bm \pi_1',\ldots,\bm \pi_T')'$. That is, where
the joint distribution function is $F_i(\bm \pi)=K_i(\bm u_i)$, with $K_i$
a $Tr$-dimensional
copula function; for example, see Nelsen~(2006; p.43). The copula function is
evaluated at the transformed data vector 
$\bm{u}_i=(\bm{u}_{i,1}',\ldots,\bm{u}_{i,T}')'$, where
$\bm{u}_{i,t}=(u_{i,1,t},\ldots,u_{i,r,t})'$,
$u_{i,j,t}=F^{\epsilon}_{i,j}(\pi_{j,t}-\eta_{i,j,t})$ is uniformly 
distributed on $[0,1]$, and $\eta_{i,j,t}=S_i(b_{i,t})+\sum_{a \in E_{i,j}}c_a(v_{a,t})$.
The values $u_{i,j,t}$ are often 
called
probability integral transformed (PIT) observations in the time 
series literature (Rosenblatt~1952),
or
`copula data' in the copula literature
when computed using estimates of $F^{\epsilon}_{i,j}$ and $\eta_{i,j,t}$. We 
later
show in Section~\ref{sec:cmts} how such copula data can be used to estimate
the copula function $K_i$.
The density function of $\bm \pi$ is obtained by differentiating its 
distribution function as
$f_i(\bm \pi) = \frac{\partial F_i(\bm{\pi})}{\partial \bm{\pi}} =
k_i(\bm{u}_i)\prod_{t=1}^T\prod_{j=1}^r f^{\epsilon}_{i,j}(\pi_{j,t}-\eta_{i,j,t})\,.$
Here,
$k_i(\bm{u}) = \frac{\partial}{\partial \bm{u}}K_i(\bm{u})$
is also a density function on $[0,1]^{Tr}$,
commonly called the `copula density'. Each
marginal density is a mixture of three components
$f^{\epsilon}_{i,j}(\epsilon)=\frac{\partial}{\partial \epsilon}F^{\epsilon}_{i,j}(\epsilon)=
\sum_{l=1}^3\omega_{i,j,l} g_l(\epsilon;\alpha_{i,j,l},\sigma_{i,j,l})$, with
$g_l=\frac{\partial}{\partial \epsilon}G_l$.


If $\Phi(x)$ denotes the standard normal distribution function, then a
Gaussian copula has the density
\begin{equation}
k_{\mbox{{\tiny Ga}}}(\bm{\xi};\Omega)=|\Omega|^{-1/2}
\exp\left\{-\frac{1}{2} \bm{w}'(\Omega^{-1}-I_n)\bm{w}
\right\}\,,
\label{eq:gcoppdf}
\end{equation}
for a vector $\bm{\xi}=(\xi_1,\ldots,\xi_n)$
with $\bm{w}=(\Phi^{-1}(\xi_1),\ldots,\Phi^{-1}(\xi_n))'$, and
a correlation matrix $\Omega$ as dependence parameters.
It is easy to show
that $\bm{w}$ is distributed
$N(0,\Omega)$, and that all sub-vectors
of $\bm{\xi}$ also have Gaussian copula densities;
for example, see Song~(2000). Here, we
assume that
$\Omega$ corresponds to the $n=Tr$ dimensional
correlation matrix of a stationary multivariate
time series $\{\bm{w}_t\}_{t=1}^T$.
In particular, we follow
Biller~(2009) and Smith \& Vahey~(2016) and employ a stationary
Gaussian VAR($p$)
for the latent process. 
The matrix $\Omega$ is a block Toeplitz 
autocorrelation
matrix, with $(s,t)$th block $R(s-t)=\mbox{Corr}(\bm{w}_s,
\bm{w}_t)$
(L\"{u}tkepohl~2006; p.30). The parameter matrix $R(0)$ captures
cross-sectional linear correlation in the latent time series process, and $R(h)$
captures serial linear correlation at lags $1\leq h\leq T-1$. However, when
combined with highly non-Gaussian margins, 
the copula framework also allows for nonlinear dependence
in the price series.

When employing a
Gaussian copula we set the copula density 
$k_i(\bm{u}_i)=k_{\mbox{{\tiny Ga}}}(\bm{u}_i;\Omega_i)$. In this case,
it is possible to show that the
multivariate series $\{\bm{u}_{i,t}\}_{t=1}^T$ of the PIT values is a
(strongly) stationary time series on the unit cube with Markov order $p$ 
(Smith~2015).
Similarly, 
the series $\{\bm{\epsilon}_{i,t}\}_{t=1}^T$ is also 
strongly stationary, where $\bm{\epsilon}_{i,t}=\bm{\pi}_t-\bm{\eta}_{i,t}$
and $\bm{\eta}_{i,t}=(\eta_{i,1,t},\ldots,\eta_{i,r,t})'$.
Measures of dependence for this time
series can be computed from the matrices $R(h)$ of the Gaussian copula,
and forecast
distributions constructed via simulation.
We discuss computation of these, along with estimation
the Gaussian copula
model, further in Section~\ref{sec:cmts}.

\section{Australian National Electricity Market} \label{sec:nem}
\vspace{-10pt} 
\subsection{The Market}
The Australian National Electricity Market
began operating in December 1998, although
regional markets were adopted several years earlier. It consists
of $r=5$ regions, which coincide with the five adjacent Australian
states of New South Wales (NSW), Victoria (VIC), Queensland (QLD), 
South Australia (SA) and Tasmania (TAS); with the latter region
joining in May 2005. All sales of electricity go through a wholesale pool, which 
is managed by the Australian Electricity Management Organization (AEMO). 
While
some limited generation and consumption of electricity does occur independently
at remote locations disconnected from the transmission grid, there are no
bilateral trades utilizing the grid. In 2011 there were 305 registered 
generators (Australian Energy Regulator 2011). While generation capacity 
based in VIC and SA 
was almost entirely privately owned during this period, approximately 90\% and 70\% of capacity in NSW
and QLD, respectively, was owned by public corporations; along with all capacity 
in TAS. Coal and gas fired generators made up the vast bulk of available capacity, with 
some hydroelectric capacity in TAS and on the NSW/VIC border, and a minor
wind and solar capacity located mainly in SA. Table~\ref{tab:capacity} provides a
breakdown of the registered capacity during 2011 by region and source. 

AEMO operates a separate price setting mechanism in each region.
Generators bid for the supply of electricity into the pool one day ahead.
Each bid consists
of 48 half-hourly prices in Australian dollars per megawatt hour (\$/MWh) 
and quantities in megawatt hours (MWh). Bids are ordered by price, and the
highest marginal
price of generation capacity dispatched (i.e. employed to meet demand) is computed
for every five minute interval. The half-hourly spot price is 
equal to the average of these six
five minute prices, and is the price at which all sales are made during that half-hour.
Re-bidding of amount (but not price) is allowed up to five 
minutes before dispatch, and is widely practised. Prior to 1 July 2010 there was a price
cap of \$10,000 per MWh, after which it was increased to \$12,500. There is a floor
price of -\$1,000. Negative prices 
occur in the NEM for short periods of time, either as a result of the high cost of ramping
down baseline coal generators, or due to strategic bidding behavior where
participants are able to exploit transmission constraints (AER 2011).
For an
introduction to the price setting mechanism see AEMO~(2010).

\begin{figure}[htbp]
\begin{center}
\epsfig{file=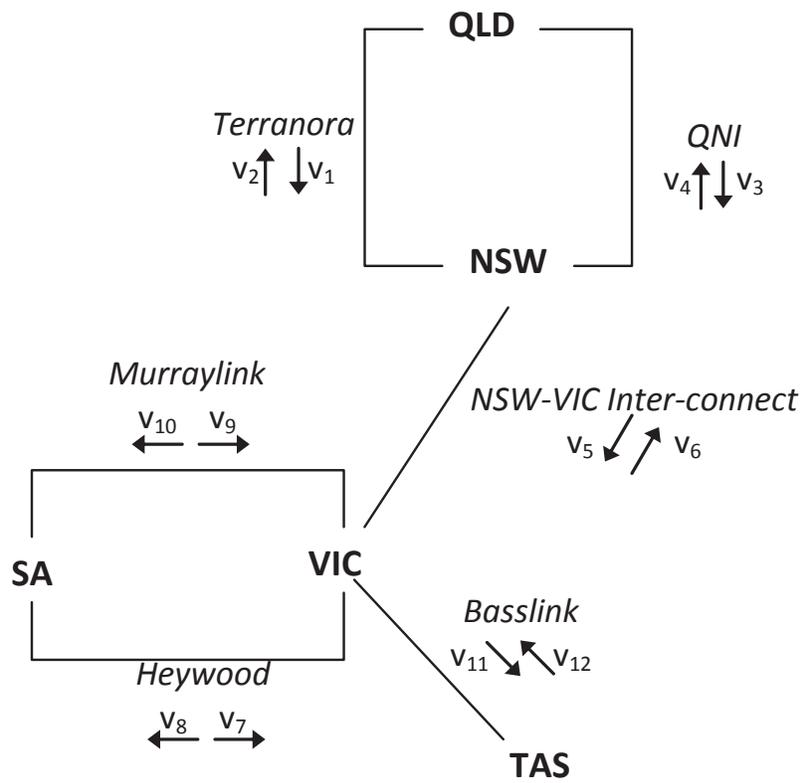,width=300pt}
\end{center}
\caption{The interconnectors in the Australian NEM, 
along with the associated directional flow variables.}
\label{fig:network}
\end{figure}

There is extensive inter-regional trade using six high voltage interconnectors. 
Figure~\ref{fig:network}
shows these, and the twelve corresponding directional flow variables,
which have
O/D pairs as listed in Table~\ref{tab:va}. From these the
sets of directed arcs for this network can be derived; for example, 
$E_{\mbox{\tiny NSW,QLD}}=\{2,4\}$ and $E_{\mbox{\tiny NSW,TAS}}=\emptyset$.
Note that of
the twenty sets of directed arcs between the five regions, twelve are empty
sets.
AEMO transmits
electricity between regions, within interconnector 
capacity constraints, with the objective of
equalizing prices up to the cost of transmission. 
When the interconnectors are at
capacity, prices in each region often differ substantially.

\subsection{The Data}
To estimate our statistical model we employ $T=17,808$ half-hourly observations on
regional loads, spot prices, and interconnector flows and losses, made
between 7 February 2010 and 13
February 2011. We use the publicly available AEMO dispatch data\footnote{We are careful to employ the dispatch, 
rather than pre-dispatch, data because it is from this that market prices are computed.}
 from the  website {\tt www.aemo.com.au}.
Data on the supply variable is computed for region $i$ from the relationship
\begin{equation}
b_i = \mbox{ load in region }i + x_i-m_i+
\mbox{interconnector loss adjustment}\,.
\label{eq:db}
\end{equation}
Here, $x_i=\sum_{j\neq i}\sum_{a \in E_{i,j}}v_a$ are total exports
from region $i$ to other regions,
$m_i=\sum_{j\neq i}\sum_{a \in E_{j,i}}v_a$ are total imports into region $i$.
For example, using the labels for interconnector flows given 
in Table~\ref{tab:va}, 
exports from NSW are $x_{\mbox{\tiny NSW}}=v_2+v_4+v_5$ and
imports into NSW are $m_{\mbox{\tiny NSW}}=
v_1+v_3+v_6$. Load in each region is the total demand for electricity, and includes any distribution
losses 
within the region.\footnote{Our load variable is labelled `TotalDemand' in the
dispatch dataset. This does not include any `normally-off' local scheduled loads, which are often exactly 
zero and are typically excluded from the definition of demand by AMEO. }
The interconnector loss adjustment arises from inter-regional transmission line losses.\footnote{Our loss adjustment
variable for each region is the `Allocated Interconnector Losses' for each region in the dispatch dataset.} 

\begin{sidewaysfigure}[htbp]
\begin{center}
\epsfig{file=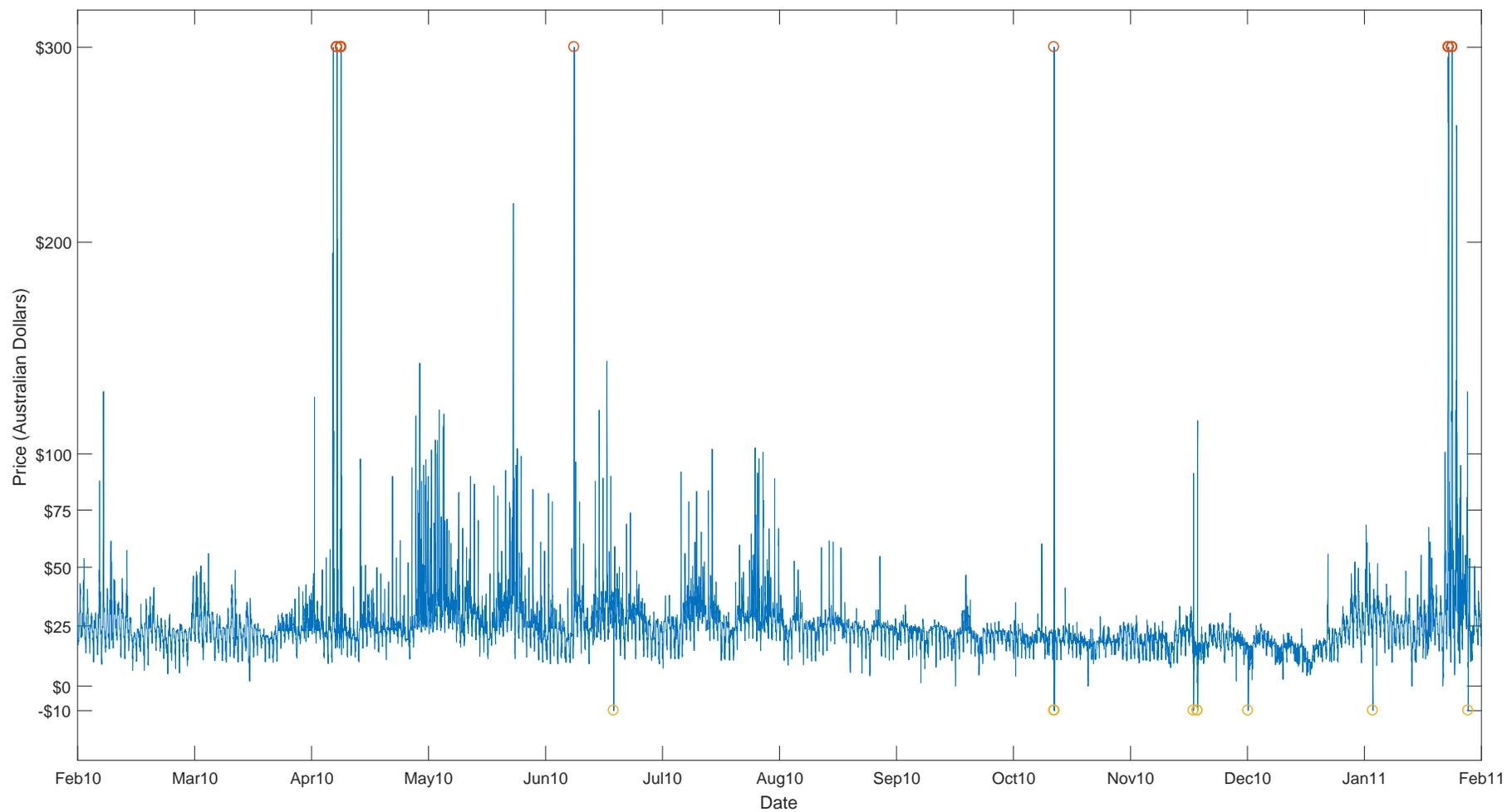,width=680pt}
\end{center}
\caption{Time series plot of half-hourly prices
for the region VIC.
Only prices in the range -\$10 and \$300 have been
plotted to
aid presentation by limiting the vertical scale. Observations
outside this range are
denoted by circles on the plot.}
\label{fig:vicTS}
\end{sidewaysfigure}

\begin{figure}[htbp]
\begin{center}
\epsfig{file=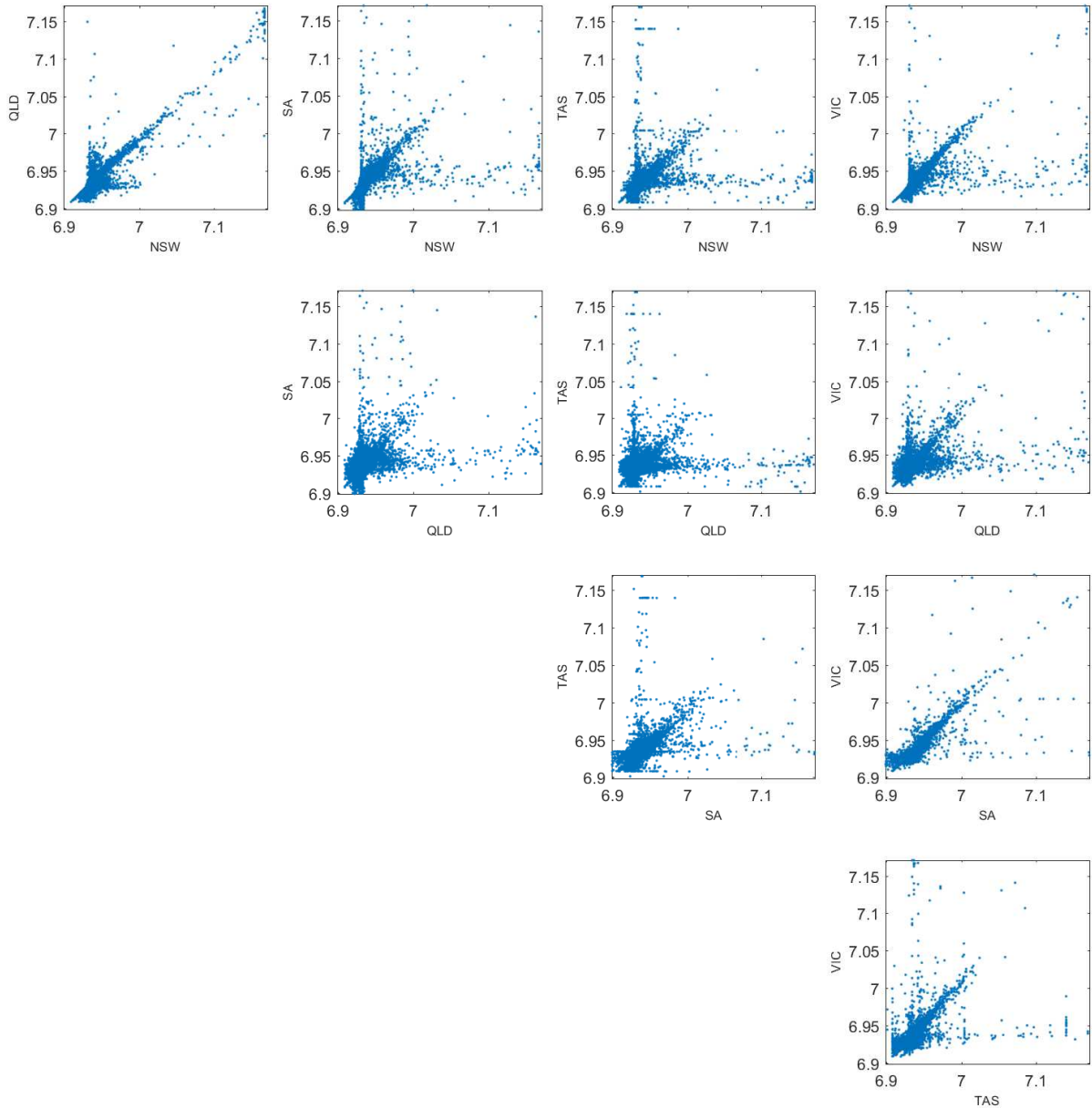,width=450pt}
\end{center}
\caption{Pairwise scatterplots of the logarithm of half-hourly prices (plus \$1001)
in the five regions of the NEM. The axes are limited to prices in the range
-\$10 and \$300 to aid presentation (ie. from $\log(991)$ to $\log(1301)$).}
\label{fig:scatter}
\end{figure}

Table~\ref{tab:pib} provides summaries of half-hourly 
price and supply for the five regions, broken
down into periods of peak (09:00--20:00) and
off-peak (20:30--08:30) demand. Mean prices during peak periods
are approximately double those during off-peak periods, except for prices
in TAS where the price difference is less. This is likely because TAS
is the only region where
supply is dominated by hydroelectric capacity, which often acts to smooth
prices. Price spikes
occur during periods of transmission
congestion and peak demand. However, inaccurate demand forecasts, unanticipated
outages, and strategic bidding can 
also cause prices to spike or fall heavily into negative territory. To illustrate,
Figure~\ref{fig:vicTS} plots
VIC regional prices on a logarithm scale. 
There were 32 observations of 
extreme prices above 500\$/MWh, and 14 of negative prices.
The time series modeling of these, and other 
stylized empirical features,
has been discussed widely in the economics,
engineering and forecasting literatures; for example,
see Karakatsani \& Bunn~(2008b), 
Panagiotelis \& Smith~(2008),
Gonz\'alez et al.~(2012), Weron~(2014), Manner et al.~(2016) 
and references therein.
Figure~\ref{fig:scatter} provides pairwise scatterplots of the five price
series on the logarithmic scale. While prices are positively dependent,
there are frequent substantial deviations between prices in different
regions at all price levels.

Table~\ref{tab:va} reports the mean and 
maximum half-hourly directed energy transmissions 
for the six interconnectors.
Both QLD and VIC
are major exporters of electricity, while
NSW and SA are predominantly 
importers. 
The nominal capacity
constraints of the directed flows are also given, although actual 
constraints vary with direction of flow 
and over time due to variation in local network thermal ratings and
voltage/reactive power limits.

\section{Estimation}\label{sec:em}
Aggregate supply curves $S_i$ are typically difficult to construct directly from
publicly available data, including for the Australian NEM. 
Moreover, the cost functions
$c_a$ are unknown in general because 
they are a feature of the econometric model that accounts for multiple aspects of the
cost of inter-regional transmission. 
Therefore, we estimate both the supply and
cost functions in Equation~(\ref{eq:stat1}) using the Bayesian
monotonic semiparametric regression methodology of Shively et al.~(2009).
These authors
demonstrate the efficiency of their approach for
Gaussian
disturbances, and we extend it here to the case where the disturbances
are distributed as a mixture of three Gaussians.
The fitted regressions
form marginal models, conditional on which the Gaussian copula model is 
estimated
using maximum likelihood. Such a two-stage approach
is used widely in
the copula literature and
is only slightly less efficient
than full maximum likelihood (Joe~2005), yet much
simpler to compute.
We outline both estimation stages 
in the following two sub-sections, although we discuss the Markov chain Monte Carlo
sampling scheme used in the first estimation stage in the Appendix.

\subsection{Monotonic Regression Models}\label{sec:mfe}

We first consider the specification of the function $S_i(b_{i,t})$ in 
Equation~(\ref{eq:stat1}).
Without loss of generality, we normalize observations
$b_{i,t}$ on the covariate to $[0,1]$, and then
approximate the function
using the quadratic regression spline
\begin{equation}
S_i^{(m)}(b_{i,t}) = \beta_{i,1}b_{i,t} + \beta_{i,2}b_{i,t}^2 +
 \beta_{i,3}(b_{i,t}-\tilde b_{i,1})_+^2 + \cdots + 
\beta_{i,m+2}(b_{i,t}-\tilde b_{i,m})_+^2\,.\label{eq:spline}
\end{equation}
Here, $\tilde b_{i,1}, \ldots, \tilde b_{i,m}$ are $m$ fixed `knots' placed 
along the domain of the independent variable $b_i$, 
such that $0 < \tilde b_{i,1} < \ldots < \tilde b_{i,m} < 1$ 
and $(z)_+=\max(0,z)$.
We set $m=25$, which is large enough to allow for a high degree of flexibility.
However, unrestricted estimation of the coefficients results in a function 
estimate that has high local variance; ie. is non-smooth. We therefore follow
the popular approach of placing a point mass probability at zero
on the coefficients, and estimate the function as a Bayesian model average;
e.g. see Smith \& Kohn~(1996).
We define $J_{i,j}=0$ if $\beta_{i,j}=0$,
and $J_{i,j}=1$ if $\beta_{i,j} \ne 0$, and assume these values are a priori 
independent with $\mbox{Pr}(J_{i,j} = 0) = p$, for $j=1,\ldots, m+2$. 
We set $p=0.8$ in our empirical work, but note that the results are insensitive to a
wide range of values. Given these priors, the final function estimate is 
obtained through model averaging over 
$\bm{J}=(J_{i,1}, \ldots, J_{i,m+2})'$. 

Monotonicity of $S_i^{(m)}(b_{i,t})$ is ensured using linear
constraints on the $\beta_{i,j}$ coefficients imposed through the prior.
Following Shively et al.~(2009), 
let $\bm{\beta}_J$ consist of the elements of
$\bm{\beta}_i=(\beta_{i,1}, \ldots, \beta_{i,m+2})'$ corresponding to those 
elements of $\bm{J}$ that are nonzero. 
Then linear restrictions on the elements of $\bm{\beta}_J$
required to impose monotonicity can be written as $L_J\bm{\beta}_J \ge \bm{0}$, 
where $L_J$ is a lower triangular matrix that depends on $\bm{J}$ and the knots,
and $L_J\bm{\beta}_J \ge \bm{0}$ means each element of the vector
is non-negative. For example, if $\bm{J}=(1,1,1,0,\ldots,0)'$,
then the three linear constraints
$\beta_{i,1} \ge 0$, $\beta_{i,1}+2\tilde b_{i,1}\beta_{i,2} \ge 0$ and 
$\beta_{i,1} + 2\beta_{i,2}+2(1-\tilde b_{i,1})\beta_{i,3} \ge 0$ ensure that $S_i^{(m)}$ is monotonically increasing.
Given $\bm{J}$ and $\sigma_{i,j,l}^2$, the prior for $\bm{\beta}_J$ is a $N(0,c\sigma_{i,j,l}^2\Omega_J)$ distribution,
constrained to the region $L_J\bm{\beta}_J \ge \bm{0}$.
Following Shively et al.~(2009), we set $\Omega_J=L_J^{-1}(L_J')^{-1}$ and $c=n$.
The functions $c_a$, $a\in E$, in Equation~(\ref{eq:stat1}) are
modeled similarly.

To account for the mixture distribution for $\epsilon_{i,j,t}$ in Equation~(\ref{eq:stat1}),
we follow the common Bayesian approach (Fr\"uhwirth-Schnatter~2006) 
of introducing latent indicators, where
$M_{i,j,t}=l$ if observation $t$ for regression $(i,j)$ is from mixture component $l$. 
Thus, $\omega_{i,j,l}=\mbox{Pr}(M_{i,j,t}=l)$, and 
$\mbox{Pr}(M_{i,j,t}=l|\bm{\pi})$ is the posterior probability that
observation $t$ is from component $l$ for regression $(i,j)$. To simplify the development of the 
sampling scheme in the Appendix, we assume each component
$G_l(\epsilon;\alpha_{i,j,l}, \sigma_{i,j,l})$
in Equation~(\ref{eq:mixdist}) is Gaussian, 
so that the marginal distribution of $\epsilon_{i,j,t}$ is a mixture of 
three Gaussians. Although we note that when computing the copula
data, we later employ nonparametric estimates of the marginal distributions.

For the mixture weights we assume a Dirichlet prior, where
$\omega_{i,j}=(\omega_{i,j,1},\omega_{i,j,2},\omega_{i,j,3}) 
\sim \mbox{Dirichlet}(1,1,1)$, so that {\em a priori}
$E(\omega_{i,j,l})=1/3$ and $\mbox{Var}(\omega_{i,j,l})=1/8$.
The component means $\alpha_{i,j}=(\alpha_{i,j,1},\alpha_{i,j,2},\alpha_{i,j,3}) \sim N(0,cI_3)$, 
constrained to the region $\alpha_{i,j,2} < \alpha_{i,j,1} < \alpha_{i,j,3}$ as 
discussed
in Section~\ref{sec:smrp}.
We set $c=100^2$, which is uninformative relative to the scale of the data.
Last, the
component variances
 $\sigma_{i,j}^2=(\sigma_{i,j,1}^2,\sigma_{i,j,2}^2,\sigma_{i,j,3}^2)$
have a uniform prior distribution on $(0,c]^3$, constrained so that
$\sigma_{i,j,1}^2<\sigma_{i,j,2}^2$ and $\sigma_{i,j,1}^2<\sigma_{i,j,3}^2$
as in Section~\ref{sec:smrp}, 
and with $c=100$.

The posterior mean $E(S_i^{(m)}(b_{i,t})|\bm{\pi})$ is used as the point estimate of $S_i(b_{i,t})$,
along with similar estimates for the functions $c_a$, $a\in E$, all of which are computed using 
the Monte Carlo iterates from the posterior. Posterior means are also used as point estimates
for the 
parameters $\omega_{i,j,l}$, $\alpha_{i,j,l}$ and $\sigma_{i,j,l}^2$.
 
\subsection{Copula Model}\label{sec:cmts}
Estimation of the copula model uses the copula data.
To compute these values, 
we first employ the marginal models fitted with the assumption that the
disturbances follow a mixture of three normals, and set 
\begin{eqnarray*}
\tilde u_{i,j,t} & = &F_{i,j}^\dagger (\hat \epsilon_{i,j,t}) \equiv 
\sum_{l=1}^3\hat \omega_{i,j,l} 
\Phi(\hat \epsilon_{i,j,t};\hat \alpha_{i,j,l},\hat \sigma_{i,j,l})\,,\mbox{ where}\\
\hat \epsilon_{i,j,t} &= 
&\pi_{j,t}-\hat \eta_{i,j,t}=
\pi_{j,t}-\hat S_i(b_{i,t}) -\sum_{a \in E_{i,j}}
\hat c_a(v_{a})\,,
\end{eqnarray*}
where parameter values with hats denote posterior mean point estimates.
In our empirical
work we find that for some marginal models $(i,j)$, the values
$\{\tilde u_{i,j,1},\ldots,\tilde u_{i,j,T}\}$ are close to uniformly distributed,
suggesting a mixture of three normals provides a good fit to the regression disturbances. However,
for other regressions they deviate meaningfully from a uniform distribution, so that to use these values
as copula data would provide poor estimates of the copula parameters. Therefore, we
construct
the empirical distribution function $F^{\mbox{\tiny EDF}}_{i,j}$ from
the values $\{\tilde u_{i,j,1},\ldots,\tilde u_{i,j,T}\}$, and compute
the copula data as 
$\hat u_{i,j,t}=F^{\mbox{\tiny EDF}}_{i,j}(\tilde u_{i,j,t})$.
For each supply region $i$,
$\hat F^{\tiny \epsilon}_{i,j}(\epsilon)=F_{i,j}^{\mbox{\tiny EDF}}
\circ F_{i,j}^\dagger(\epsilon)$ 
provides a nonparametric estimate of the marginal distribution function
in Equation~(\ref{eq:mixdist}).
The approach of employing nonparametric
estimates for marginal distributions, followed by a parametric copula, is 
popular in multivariate modeling; for example, see Shih \& Louis~(1995).

One advantage of using a Gaussian copula
for $K_i$ is that, conditional on the copula data,
estimation of the copula parameter matrix $\Omega_i$
by maximum likelihood is straightforward using the VAR representation. Let 
$\hat w_{i,j,t}=\Phi^{-1}(\hat u_{i,j,t})$ and
$\hat{\bm{w}}_{i,t}=(\hat w_{i,1,t},\ldots,\hat w_{i,r,t})'$, then
the MLE of
the autoregressive parameters of a stationary VAR($p$)
for the time series $\{\hat{\bm{w}}_{i,t}\}_{t=1}^T$ are readily
computed;
for example,
see Mauricio~(1995).\footnote{The stable 
VAR models for the series $\{\bm{\hat w}_{i,t}\}_{t=1}^T$
are estimated in Matlab using the routine `vgxvar'.} 
From these, point estimates of the autocovariance matrices
$\Gamma_i(h)=\mbox{Cov}(\bm{\hat w}_{i,t},\bm{\hat w}_{i,t-h})$ can be computed as in L\"utkepohl~(2006; pp.28-31), along with estimates of
the
corresponding autocorrelation matrices $R_i(h)=D_i(h)^{-1/2}\Gamma_i(h)D_i(h)^{-1/2}$, where the 
diagonal matrix $D_i(h)=\mbox{diag}(\Gamma_i(h))$. The matrices $R_i(h)$ 
are the component blocks
of $\Omega_i$, and the resulting estimate is its MLE; for example,
see Song~(2000). We note here that Smith \& Vahey~(2016) propose an alternative Bayesian
estimator for the copula parameters of this model based on a drawable vine representation
of the Gaussian copula density (Bedford and Cooke~2002) 
when the latent Gaussian process has Markov
order 4. However, in our empirical analysis of half-hourly electricity prices, 
a long Markov order of $p=7\times 48=336$ is used,
so that a vine decomposition
involves far too many pair-copula terms to be evaluated in practice.
In comparison, computation of the MLE of this copula time series model using the
VAR representation is computationally feasible using existing software.

Another advantage of using the Gaussian copula is that
it is straightforward to obtain inference on the mean-corrected
time series
$\{\bm{\epsilon}_{i,t}\}_{t=1}^T$. This includes measures of
dependence
and the predictive distributions, as we now
outline.
If $\phi_{j,l}(h)$ is the $j,l$-th element of $R_i(h)$,
then the pairwise dependence between $\epsilon_{i,j,t}$ and $\epsilon_{i,l,t-h}$,
can be measured using Kendall's tau, which 
for a Gaussian copula is
$\tau_{j,l}(h)=\frac{6}{\pi}\mbox{arcsin}\left(\frac{\phi_{j,l}(h)}{2}\right)$.
These values can be arranged into $(r\times r)$ matrices
${\cal T}(h)$, which have 
$j,l$-th elements $\tau_{j,l}(h)$. Then,
${\cal T}(0)$
measures cross-sectional dependence of the vector $\bm{\epsilon}_{i,t}$, and
${\cal T}(h)$ measures serial dependence at lags $h\geq 1$. 
We note that ${\cal T}(h)$ is symmetric for $h=0$, but
asymmetric for $h\ge 1$. We call these matrices `auto-dependence'
matrices, as they are direct generalizations of the autocorrelation
matrices in linear multivariate time series, and we compute some examples
in our later empirical work.
Similar matrices
can also be constructed
from other pairwise measures of dependence, such as Spearman correlations.

The $h$-step ahead predictive distribution with conditional density
\begin{equation}
f(\bm{\epsilon}_{i,T+h}|\bm{\hat \epsilon}_{i,T-p},\ldots,\bm{\hat \epsilon}_{i,T})\,,
\label{eq:cdist}
\end{equation}
can be evaluated
in a Monte Carlo fashion as follows. Iterates
of the latent time series $h$-steps ahead $\{\bm{w}_{i,T+1},\ldots,\bm{w}_{i,T+h}\}$ 
are generated from
the fitted VAR($p$) model,\footnote{This is undertaken in Matlab using the routine `vgxsim'.}
conditioning on the previous values 
$\{\bm{ \hat w}_{i,T-p},\ldots,\bm{\hat w}_{i,T}\}$ (which are computed
from $\{\bm{\hat \epsilon}_{i,T-p},\ldots,\bm{\hat \epsilon}_{i,T}\}$ as discussed above).
The elements of each iterate $\bm{w}_{i,T+h}$ are then transformed as
$u_{i,j,T+h}=\Phi(w_{i,j,T+h})$, and
further
transformed as $\epsilon_{i,j,T+h}=\left(\hat{F}^\epsilon_{i,j}\right)^{-1}(u_{i,j,T+h})$.
These successive transformations produce
iterates of $\bm{\epsilon}_{i,T+h}$ that are distributed with the
density given above in Equation~(\ref{eq:cdist}).

\section{Density Forecasting}\label{sec:dfore}
Predictive distributions of prices
are constructed from the statistical model in two
ways. The first is conditional on observed values of
supply and interconnector flows. However, while AMEO is likely to have accurate
forecasts for these values, they will not be known to market
participants in general.
Therefore, we also outline a
second approach
where supply and interconnector
flows are also forecast.  In either case,
because there
is a separate statistical model for each supply region $i$,
there are $r$ separate
forecast
distributions. We then combine these distributions
to produce an ensemble forecast distribution. 

\subsection{Conditional on supply and interconnector flows}\label{sec:cons}
We first consider the case when supply and interconnector flows
are assumed known
during the forecast window.
Let $\bm{v}_t$ be a vector of the flows along all interconnectors at time $t$.
Then for supply region $i$, 
the
predictive density of the price vector $\bm{\pi}_t=(\pi_{1,t},\ldots,\pi_{r,t})'$
$h$ steps ahead from
forecast origin $T$ is
\begin{equation}
f^{(i)}(\bm{\pi}_{T+h}|{\cal F}_T,{\bm v}_{T+h},b_{i,T+h})
=\bm{\eta}_{i,T+h}+f(\bm{\epsilon}_{i,T+h}|\bm{\epsilon}_{i,T-p},
\ldots,\bm{\epsilon}_{i,T})\,.
\label{eq:predi}
\end{equation}
This forecast is conditional on information at time $T$ (ie. the filtration ${\cal F}_T$)
and also the future
supply 
$b_{i,T+h}$ and 
interconnector flows
$\bm{v}_{T+h}$.
In Equation~(\ref{eq:predi}) the 
value
$\bm{\eta}_{i,T+h}$ can be
computed from the fitted marginal models by plugging in 
$b_{i,T+h}$ and $\bm{v}_{T+h}$, while
the predictive distribution 
of $\bm{\epsilon}_{i,T+h}$ 
can be
computed via simulation from the fitted copula time series model as
discussed in Section~\ref{sec:cmts}.

Because there is a separate predictive density for each supply
region $i$, we combine these into an 
ensemble predictive distribution with density
\begin{equation}
f^{\mbox{\tiny Ens}}(\bm{\pi}_{T+h})=
\sum_{i=1}^r W_i \left[
f^{(i)}(\bm{\pi}_{T+h}|{\cal F}_T,{\bm v}_{T+h},b_{i,T+h})
\right] \,.
\label{eq:ens1}
\end{equation}
The weights $\{W_1,\ldots,W_r\}$ 
should not be confused with the weights
of the mixture distribution at Equation~(\ref{eq:mixdist}). They
are assumed equally-valued (ie. $W_i=1/r$) in our empirical work
for simplicity, although other approaches for determining these
can also be employed (Jore et al.~2010). Each component in the ensemble
has predictive mean
\[
\bm{\pi}^{(i)}_{T+h|T}=\bm{\eta}_{i,T+h}+
E(\bm{\epsilon}_{i,T+h}|\bm{\epsilon}_{i,T-p},\ldots,\bm{\epsilon}_{i,T})\,,
\]
which can be computed from the fitted model.
The weighted sum
$\bm{\pi}_{T+h|T}^{\mbox{\tiny Ens}}=\sum_{i=1}^r W_i \bm{\pi}^{(i)}_{T+h|T}$
of these
predictive means
is used as a point forecast.

\subsection{Joint with supply and interconnector flows}\label{sec:jntfl}
To predict supply, we first predict load in each region at an 
intraday resolution.
Forecasting intraday
electricity load is a
well-studied problem, and there are a number of effective solutions
which can be employed here; 
see the overview
by Weron \& Misiorek~(2008).
For region $i$, at time $t>T$ we denote 
load as $d_{i,t}$, total imports 
as $m_{i,t}=\sum_{j=1}^r \sum_{a \in E_{j,i}}v_{a,t}$ and total
exports as $x_{i,t}=\sum_{j=1}^r \sum_{a\in E_{i,j}}v_{a,t}$. 
Then,
from Equation~(\ref{eq:db}), 
supply in region $i$ at time $t$ is $b_{i,t}=d_{i,t}+x_{i,t}-m_{i,t}+
\mbox{ interconnector loss adjustment}$.\footnote{For simplicity, in our 
later validation study we use actual demand and allocated transmission losses, but 
note that forecasts for both are usually readily available to system operators and can be
used instead.}

Therefore, given load forecast $d_{i,T+h}$, the problem
becomes one of forecasting the future transmission flows $\bm{v}_{T+h}$ from
which $m_{i,T+h},x_{i,T+h}$ and $b_{i,T+h}$ can be computed.
To undertake this
we solve an
optimization problem that minimizes differences between
expected prices in each region, where the expectations are obtained from our
statistical model.
For the Australian NEM, this approach mirrors the objectives of
inter-regional transmission scheduling
by the network management
organization AEMO. There is a separate optimization problem
for each component $i$ of the ensemble distribution, with an
objective function $D_{T+h}(\bm{\pi}^{(i)}_{T+h|T})$, where
\begin{equation}
D_{T+h}(a_1,\ldots,a_r)=
\sum_{j=2}^r \sum_{l<j} \delta_{l,j} |a_j-a_l|\,.\label{eq:Dfun}
\end{equation}
The values $\delta_{l,j}$
are weights, which we set as proportional to the demand in regions $l$ and $j$
at time $T+h$,
so that 
$\delta_{l,j}=(d_{l,T+h}+d_{j,T+h})/\sum_{j=2}^r\sum_{l<j}(d_{l,T+h}+d_{j,T+h})$. 
This increases the contribution of price differences between regions 
with higher
demand, and down-weights that for regions with lower demand. The
objective function is minimized with respect to 
$\bm{v}_{T+h}$, given system constraints on interconnector flows. 

For our NEM data
we incorporate two sets of system constraints. The first set of constraints
is the upper bound on the capacity of the interconnectors.
Table~\ref{tab:va} outlines nominal upper capacity on the interconnectors
in the NEM. However, actual upper capacity differs at any given 
point in time depending upon a number of factors, so that
the observed maximum flows provide more realistic estimates of
upper bounds, and we use these as the first set of constraints. The second
set of constraints derive from our construction of the directed flow variables
from net bi-directional flows along each interconnector. 
For example, $v_{5}$ and $v_6$ are
directed flows along the NSW-VIC Interconnect, with $v_5>0$ only when 
$v_6=0$ (and vice versa). This corresponds 
to the equality constraint $\max(v_5,v_6)=v_5+v_6$; similarly for the other directed
flow pairings depicted in Figure~\ref{fig:network}.

The optimization is repeated for all periods in the forecast horizon and
each supply region $i=1,\ldots,r$, producing forecast values
of interconnector flows, from which supply values are also computed.
Forecasting of prices then proceeds
as outlined in Section~\ref{sec:cons}. 

\section{Empirical Analysis}\label{sec:emp}

\subsection{Monotonic Regressions}
We employ the logarithm of spot prices as the price
data $\pi_1,\ldots,\pi_5$.
Because prices are occasionally
negative,
for each price series
the logarithm is computed
after subtracting 
the minimum pool price of -\$1000, and adding \$1. We note that 
even after a logarithmic transformation, prices are right skewed and
have a heavy right tail; which is typical of wholesale electricity prices
generally.
The Bayesian method is employed to estimate each of
the 25 regressions, and posterior
means of the monotonic functions and
model parameters are computed from the Monte Carlo samples.
 
For example, Table~\ref{tab:margins} provides the parameter
estimates of the mixture distribution components 
in each of the five VIC supply regressions. We focus
on these regression results because VIC plays a key role as the central
region
of the interconnected NEM; for example, see the recent discussion by Han, Kordzakhia \& Tr\"uck~(2017).
Component~1 has low price variation and
occurs 75\% of the time for NSW prices,
86\% of the time for QLD prices,
90\% of the time for SA prices, 90\% of the time for TAS prices,
and 93\% of the time for VIC prices.
For price formation in NSW, QLD, TAS and VIC, Component~2 captures
prices with the same mean, but with a standard deviation between 5 and 7 times
larger than Component~1. For the same four regions,
Component~3 captures the price spikes, with substantial increases in average
prices, a standard deviation between 126 and 209 times larger than Component~1,
and infrequent occurrence between 0.4\% and 2.7\% of the time.
For price formation in SA, Component~3 still has a higher average price,
but variation that is only 4 times greater than that of prices in Component~1.
Instead, Component~2 captures prices
with extreme variation and a low incidence of 0.8\%. The results suggest
that price formation in SA differs from that in other regions. Certainly,
the SA price distribution differs in
Table~\ref{tab:pib}, with more
negative prices and higher average prices than other regions. While not reported here,
interpretation of the mixture components are similar
for the regressions of
prices based on the other four supply regions.

\begin{figure}[htbp]
\begin{center}
\epsfig{file=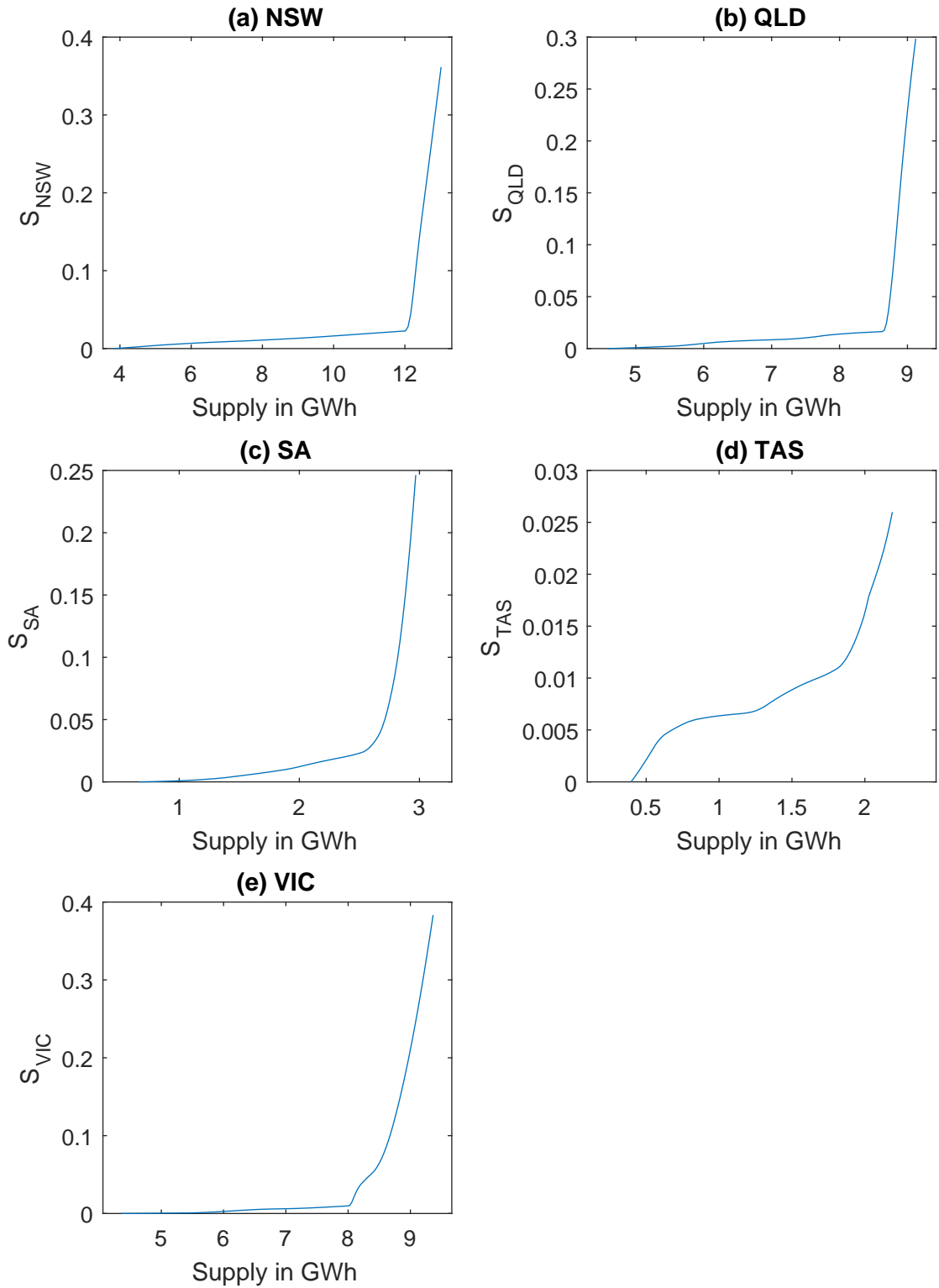,width=420pt}
\end{center}
\caption{Estimates of the five supply functions. Each function is an 
ensemble of five posterior means, where each posterior mean is
an estimate from a monotonic regression.}
\label{fig:supply}
\end{figure}

There are five monotonically increasing estimates of each (inverse) supply
function $S_i$, which we
combine into a single ensemble estimate 
$\hat S_i=\frac{1}{5}\sum_{j=1}^5 E(S_i|\pi_{j,1},\ldots,\pi_{j,T})$,
that is also monotonically
increasing. Figure~\ref{fig:supply} plots
the ensemble estimates of each supply function, and a number of 
observations can be made. The left hand side
of each function is flat, as supply is made up of low cost baseline 
capacity, and a kink and rapidly increasing right-hand side corresponds
to capacity with rapidly increasing marginal costs. 
Table~\ref{tab:capacity} provides a summary 
of registered 
generation capacity in each region at 2011. 
Coal is the lowest cost, followed by
hydroelectric, gas and lastly `other capacity' (the latter
of which includes renewable sources). 
For example, in VIC there is
8.8MW of coal and hydroelectric capacity, which is approximately where
the kink in the supply function occurs. In TAS, prices rise when
supply goes beyond around 2MW, close
to the hydroelectric capacity of 2.3MW. 

\begin{figure}[htbp]
\begin{center}
\epsfig{file=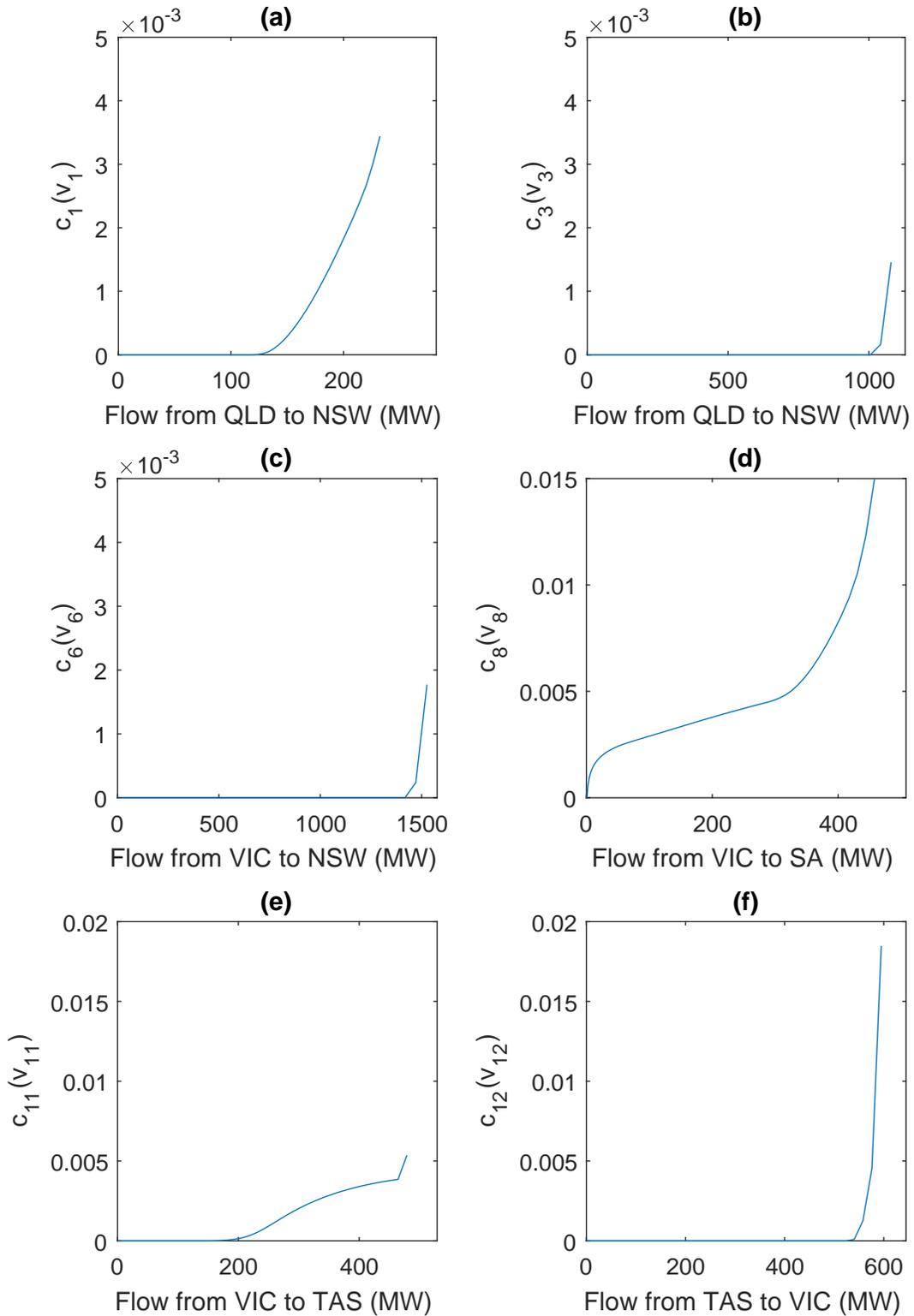,width=400pt}
\end{center}
\caption{Estimates of the cost functions for the six main interconnector
flow variables. Panels~(a) to~(f) are posterior mean estimates of
the function 
$c_1$, $c_3$, $c_6$, $c_8$, $c_{11}$ and $c_{12}$, respectively. Summaries
of the corresponding half-hourly directed transmission flows are given in Table~2.}
\label{fig:costs}
\end{figure}

Figure~\ref{fig:costs} plots the posterior mean estimates of the 
cost functions for the major inter-regional trade 
flows in Table~\ref{tab:va}. NSW is the major importer of electricity
in the NEM, 
and
panels~(a), (b) and~(c) show that imports
in excess of 150 MWh on the
Terranora, 1000 MWh on the 
QNI, and 1400 MWh on the NSW-VIC interconnect,
correspond to an increase in NSW prices. SA is the second major importer of electricity
in the NEM, with panel~(d) showing that increased flows from VIC correspond to higher
SA prices. The Basslink interconnector is the only third
party commercially operated interconnector
in the NEM, with all others being owned and operated directly by
AEMO. Imports into TAS produce only a limited impact on local prices,
but exports to VIC
in excess of around 520MWh from TAS correspond to a sharper increase in VIC prices.
Nevertheless, increased flows on the interconnectors are unrelated to substantial
regional price variation, which is consistent with a similar observation
by Higgs et al.~(2015).

\subsection{Copula Models}
The copula data are computed from the fitted marginals, and the five copula
multivariate time series models estimated using MLE for each supply region $i=1,\ldots,5$.
 We use the Bayesian information criterion (BIC) to identify
a lag structure from all thirty-five combinations of lags between 1 and 5 half-hours,
and also lags
at the same time of the day between 1 to 7 days previously (i.e. lags at 48, 96, 144, ... , 336 half-hours).
An almost
identical Markov structure was identified for all five models, with lags at
the same half-hour of the day 1,2,3 and 
7 days previously, and also at the three (TAS, VIC supply regions)
and four (NSW, QLD, SA supply regions) half-hours immediately previous. 
In all cases, 
there is substantial
serial and cross-sectional dependence in the copula data.

\begin{figure}[htbp]
\begin{center}
\epsfig{file=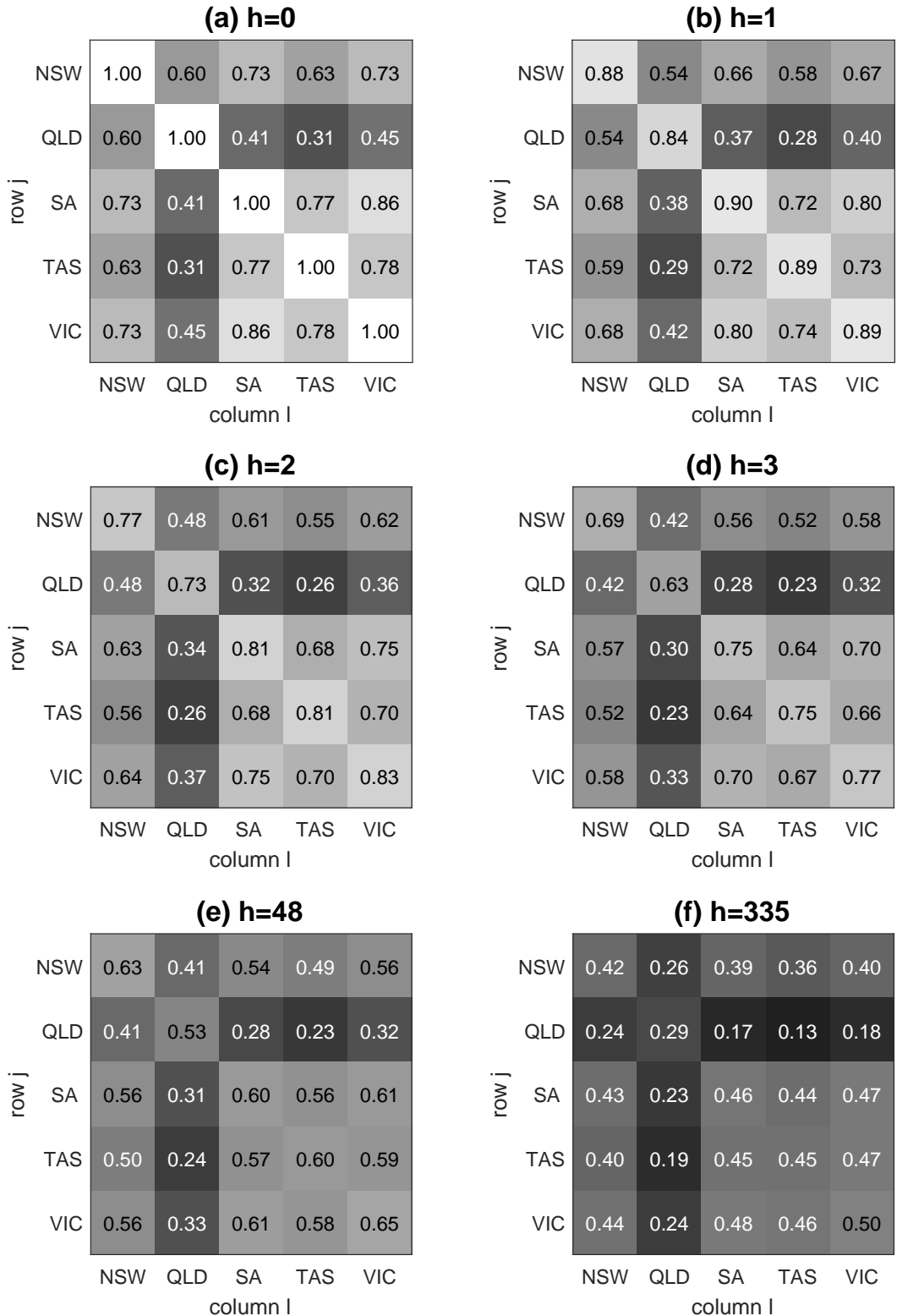,width=400pt}
\end{center}
\caption{Estimates from the VIC supply copula model of the auto-dependence matrices (a) ${\cal T}(0)$,
(b) ${\cal T}(1)$, (c) ${\cal T}(2)$, (d) ${\cal T}(3)$, (e) ${\cal T}(48)$ and (f) ${\cal T}(p-1)$, where $p=336$ corresponds to a 7 day lag length.
The matrices are defined in Section~4.2. Each $j,l$th element 
is the Kendall's tau value $\tau_{j,l}(h)$, 
which is the unconditional
pairwise  dependence
between
$\epsilon_{\mbox{\tiny VIC},j,t}$ and $\epsilon_{\mbox{\tiny VIC},l,t-h}$. 
For example,
in panel~(d) the Kendall's tau 
between $\epsilon_{\mbox{\tiny VIC},3,t}$ and $\epsilon_{\mbox{\tiny VIC},2,t-3}$
is $\tau_{3,2}(3)=0.30$, whereas the Kendall's tau between 
$\epsilon_{\mbox{\tiny VIC},2,t}$ and $\epsilon_{\mbox{\tiny VIC},3,t-3}$
is $\tau_{2,3}(3)=0.28$.}
\label{fig:auto}
\end{figure}

To summarize the overall dependence structure,
Figure~\ref{fig:auto} plots the auto-dependence matrices ${\cal T}(h)$ for lags $h=0$, 1, 2, 3, 48 and 335,
arising from the copula model with the VIC supply
regressions as margins. These are
computed from the autocovariance matrices 
$\Gamma(h)=\mbox{Cov}(\bm{w}_{\mbox{\tiny VIC},t}, \bm{w}_{\mbox{\tiny VIC},t-h})$, which are evaluated
using sparse matrix algebra applied to the VAR(1) representation
of a VAR($p$) process; see L\"{u}tkepohl~(2006; pp. 26-31). Sparse calculations are important here as this
involves algebra employing $(r^2p^2 \times r^2p^2)$ dimensional matrices with
$r=5$ and $p=336$.\footnote{This is implemented in Matlab in routines var2auto.m and plotautocorr.m 
found in the Supplementary Material.} 
These measure dependence in prices, corrected for the 
impact of the VIC supply regression marginals, and a number of observations can be made. First,
there is positive serial dependence in prices throughout. This is consistent
with previous research that finds that there is significant time
series dependence in electricity prices, even when corrected for
a number of demand
and supply-side variables (Gonz\'alez et al.~2012). Second, this serial dependence declines
as the lag $h$ increases, which is due to the assumption of stationarity in the copula
time series model. The decline is fastest for the prices in QLD, with $\tau_{2,2}(335)=0.29$, and
slowest for prices in VIC, with $\tau_{5,5}(335)=0.5$. 
Third, QLD prices are least
dependent with those in other regions. We note also that
QLD is the region with the lowest
level of average imports and also lowest prices. Last, 
VIC, SA and TAS exhibit the strongest cross-sectional dependence in prices
at all
lags, suggesting a higher level of market integration between these regions.

\subsection{Event Studies}
A key feature of the proposed econometric model is that it
can be used to measure the response in regional prices
to supply shocks or price impulses in any one region. Supply-side shocks are transmitted
to prices
through the supply function estimates, while price impulses are 
transmitted to future prices through the copula multivariate time series model. 
\subsubsection{Supply Shock}
Consider the impact of a hypothetical supply shock
in region $i$ at time $t$ by an amount $\bar b$,
equivalent to a major baseline generator being taken offline.
The marginal impact on prices can be evaluated using the $i$th supply
region regressions, but where
the supply curve $S_i$ is shifted to the left
by $\bar b$, producing a new supply curve
$\bar S_i(b_{i,t})=S_i(b_{i,t}+\bar b)$. The expected (logarithm of)
price in region $j$ from this model is then
$E(\pi_{j,t})=\bar \eta_{i,j,t}+ \tilde \alpha_{i,j}$, where
$\bar \eta_{i,j,t}=S_i(b_{i,t}+\bar b)+\sum_{a \in E_{i,j}}c_a(v_{a,t})$.
\footnote{Because we are modeling the logarithm of prices, the
expectation of actual price can be computed as the sample
mean of
exponentiated iterates simulated from the marginal regression model in 
Equation~(2.3).}

\begin{figure}[htbp]
\begin{center}
\epsfig{file=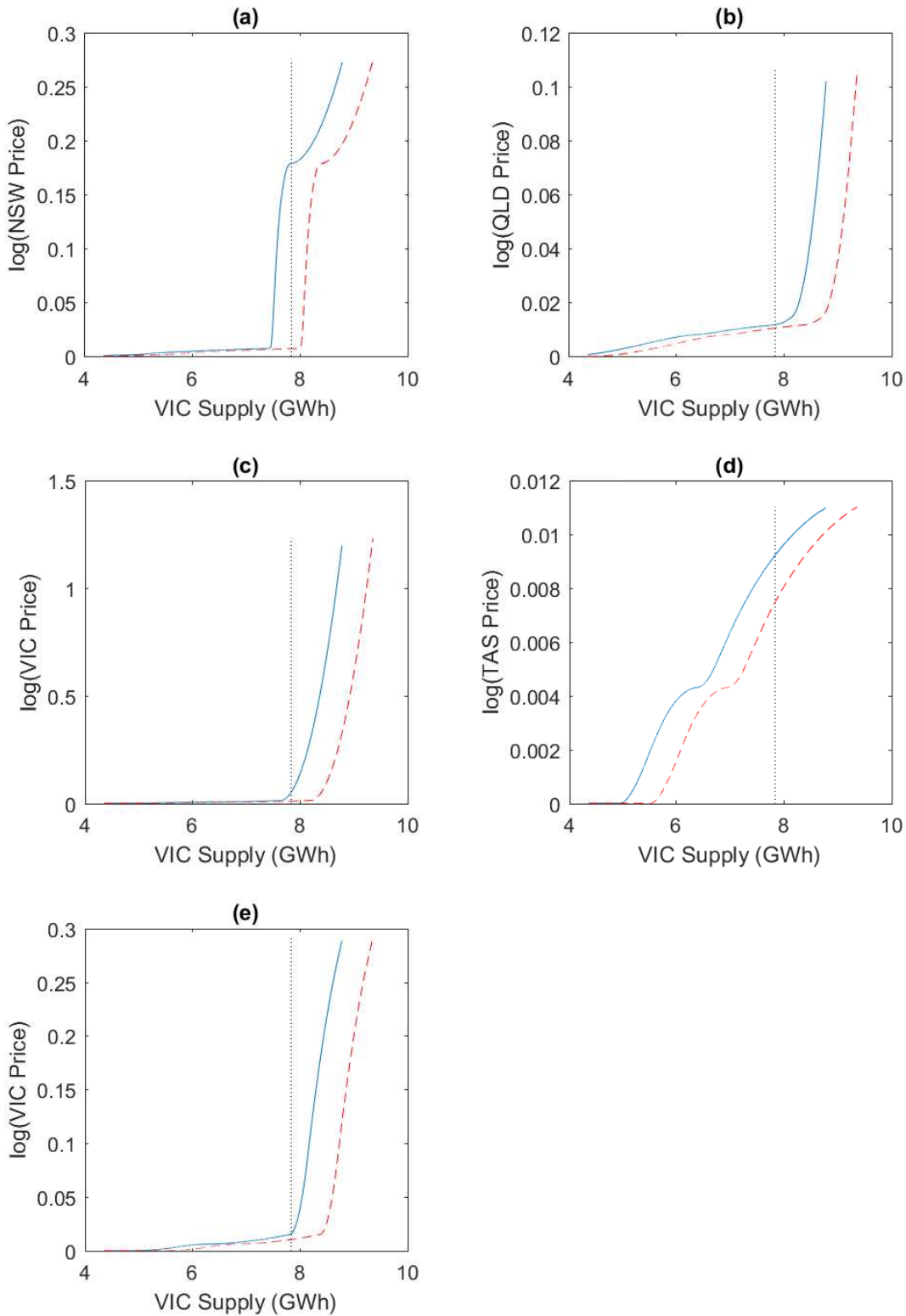,width=400pt}
\end{center}
\caption{Posterior mean estimates of the
supply curves $E(S_{\mbox{\tiny VIC}}|\pi_{j,1},\ldots,\pi_{j,T})$
(red dashed lines) from the VIC supply regressions, and
supply curves $E(\bar S_{\mbox{\tiny VIC}}|\pi_{j,1},\ldots,\pi_{j,T})$
(blue solid lines) shifted by a supply shock of 560MWh.
Panel~(a) is for NSW prices ($j=1$), (b) for QLD prices ($j=2$), 
(c)~for SA prices ($j=3$), (d)~for TAS prices ($j=4$), and~(e) for VIC 
prices ($j=5$). The vertical line marks 7845.4MWh, which is VIC supply
on 30 June 2010 at 8:00.}
\label{fig:Sbar}
\end{figure}

To illustrate, we
consider a supply shock of 560 MWh in VIC, which is 
equivalent 
to Steam Turbine Number 1 at the Loy Yang A Power Station going offline.
The Loy Yang A Power Station is the largest single registered baseline generator
in the VIC
region, and the loss of Steam Turbine Number 1 is
equivalent to the loss of approximately 4.6\% of total 2011 
registered capacity in that region. Outages of this size or larger 
occur from time-to-time in the NEM due to plant or line failure.
Figure~\ref{fig:Sbar} plots the
posterior means
$E(S_{\mbox{\tiny VIC}}|\pi_{j,1},\ldots,\pi_{j,T})$ and 
$E(\bar S_{\mbox{\tiny VIC}}|\pi_{j,1},\ldots,\pi_{j,T})$ for each of the five
price series $j=1,\ldots,5$. Using these estimates, along with the other
components of the fitted model, we compute the marginal expected prices
with and without the shock on 30 June 2010 at 08:00.
Supply during this half-hour was $b_{i,t}=7845.4$MWh, which is the 99.3th 
percentile
recorded in the data for this half-hour, and the 97.95th percentile over all half-hours,
so that VIC supply is at a high level before the shock. Therefore, the
shock results in a substantial increase in prices, with the vector of increases
in marginal expected prices
being (194.70, 1.28, 32.74, 1.85, 5.07) \$/MWh for (NSW, QLD, SA, TAS, VIC).
The increase is greatest in NSW, 
which Figure~\ref{fig:Sbar}(a) shows has an
estimated supply function $E(S_{\mbox{\tiny VIC}}|\pi_{\mbox{\tiny NSW},1},\ldots,
\pi_{\mbox{\tiny NSW},T})$ 
with a kink at the lowest VIC supply value (around 8GWh). Prices in QLD, SA and VIC
are less responsive to VIC supply shocks, with kinks occurring at higher values
of VIC supply.



\subsubsection{Price Impulse}
The
impact of price shocks not caused by changes in supply, such
as unexpected changes to demand or bidding behavior, are captured by
the multivariate time series $\{\bm{\epsilon}_{i,t}\}$.
The impact of such a shock
can be measured by comparing the predictive distribution of prices for 
subsequent periods with, and without
the shock. This is a popular approach, and 
for
nonlinear multivariate time series analysis, it is often called a generalized
impulse response analysis; for example, see
Koop, Pesaran \& Potter~(1996).
For supply region $i$, let
$\bm{\bar \epsilon}_{i,t}$ be the value of the time series at time $t$
with $j_1$th element $\bar \epsilon_{i,j_1,t}=\epsilon_{i,j_1,t}+\bar \pi$,
where $\bar \pi$ is the price shock in region $j_1$ on the logarithmic 
scale.\footnote{For example, if this is a
200 \$/MWh increase at time $t$, then the increase on the logarithmic 
scale is 
$\bar \pi=\log(\mbox{Price}_{j_1,t}+1001+200)-\log(\mbox{Price}_{j_1,t}+1001)$.}
Then we compute the forecast distribution of
$\pi_{j_2,t+h}|{\cal F}_t$ with and without the shock over a horizon of $h$ half-hours
ahead.
This is computed
by simulating Monte Carlo iterates
from  the
copula multivariate time series model 
for future values $\{\bm{\epsilon}_{t+h}\}$.

\begin{sidewaysfigure}
\begin{center}
\epsfig{file=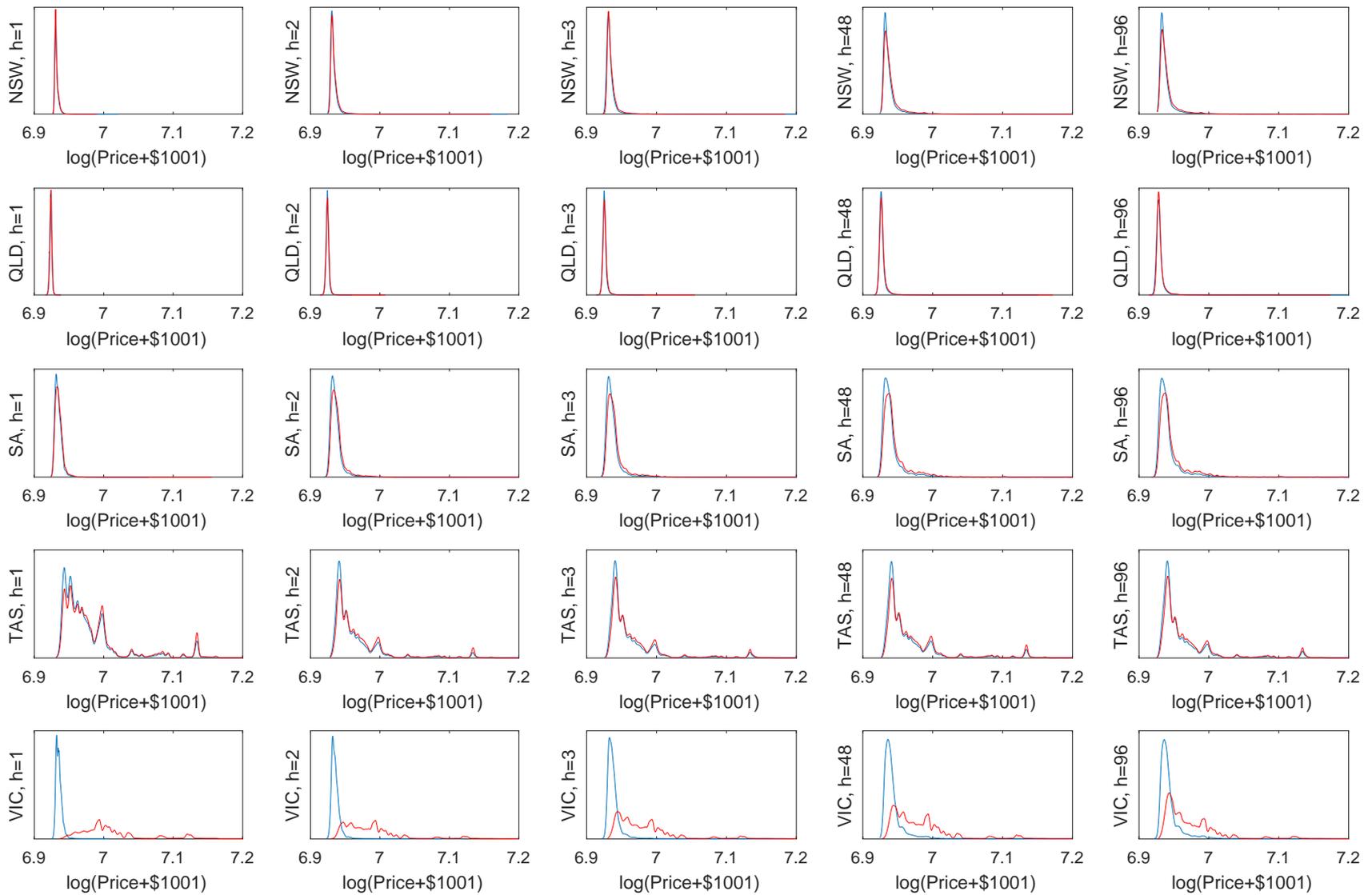,width=640pt}
\end{center}
\caption{Predictive densities of price with (red line) and without
(blue line) a price shock of \$200 in VIC during the two hour period 03:00-04:59
on 19 May 2010. The panels are arranged
as a $5\times 5$ matrix, with rows
corresponding to prices in different regions,
and columns corresponding to predictive distributions 
$h=1, 2, 3, 48$ and 96 half-hours ahead. Prices
are on the logarithmic scale for presentation,
and the density estimates are computed
using a kernel method with locally adaptive bandwidth. For presentation purposes
the horizontal axes are bounded to the right at 7.2, although all logarithmic
price
distributions have long right tails that go far beyond this 
point. While the predictive distributions can look similar for prices in some states, the price shock
can accentuate the right tail --- thereby increasing the mean --- as discussed in the text.}
\label{fig:IRF}
\end{sidewaysfigure}

To illustrate, we consider the impact of a \$200 increase in prices
(ie. an `impulse')
in VIC on prices in all regions using the fitted copula model with margins 
given by the VIC supply regressions. We consider the impulse over the 
two hour period 03:00--04:59 on 19 May 2010, denoted as half-hours $t-3,t-2,t-1$ and $t$. 
Figure~\ref{fig:IRF} plots
estimates of the predictive distributions of
$\pi_{j,t+h}$, both without the shock (blue lines)
and with the shock (red lines). 
The panels are arranged
as a $5\times 5$ matrix, with rows
corresponding to prices in different regions,
and columns corresponding to predictive distributions 
$h=1, 2, 3, 48$ and 96 half-hours ahead. Prices
are on the logarithmic scale, and the density estimates are computed
using a kernel method with locally adaptive bandwidth (Shimazaki \& Shinomoto~2010).
All predictive distributions have a long right tail, which
has been omitted here for presentation purposes. The
non-smooth nature of the presented densities is not due to either 
Monte Carlo sample error,\footnote{We use 50,000 Monte Carlo iterates to 
construct each density highly accurately.} nor due to
inefficiencies in the kernel density estimates. Instead, it is because in the marginal
models the nonparametric empirical distribution functions 
were employed as part of the model,
and these are non-smooth.\footnote{Alternatively, smooth predictive densities can be readily 
obtained by employing smooth parametric models --- such as regressions with skew t disturbances --- for the margins.}

The response to the price shock is strongest at shorter
horizons and in the same region as the price impulse (ie. in VIC). The
bottom row of panels show the large impact on the predictive distribution of price in VIC,
with the impulse increasing the predictive mean by 868.53, 438.54, 403.02, 519.33 and 317.05 \$/MWh
at horizons $h=1,2,3, 48$ and 96 half-hours ahead, respectively. 
Prices
in QLD are almost completely
unaffected by the impulse in VIC, which is consistent with the low pairwise
dependence between prices in these two regions depicted in Figure~\ref{fig:auto}.
While it initially appears in Figure~\ref{fig:IRF}
that the impulse has limited effect on prices
in NSW, SA and TAS, that is not the case. The impulse substantially
accentuates the 
right hand tail of the predictive distributions in these regions, making 
price spikes more likely in the forecast horizon and
increases the means. 
To illustrate, the price impulse in VIC increases mean predictive prices one day ($h=48$)
ahead in NSW, SA and TAS
by 5.12, 26.03 and 123.46 \$/MWh, respectively.

\subsection{Validation Study}
To validate our fitted model we undertake a
limited forecasting study. A number of different
models were fit to hourly data and
used to forecast hourly prices.\footnote{Hourly, rather than half-hourly, data are used to reduce the computational burden of the validation study. However, the models fitted to hourly and half-hourly data are very similar, so that the results are unlikely to differ meaningfully.} We construct forecasts with a daily expanding window with 100 forecast origins from 24 October 2010 to 31 January 2011. From each origin, we construct forecasts of the five regional prices over a horizon of one week at the hourly resolution.\footnote{In total, we produce forecasts for each of the five regional
prices at 
$100 \times 24 \times 7 = 16,800$ hours, so that a total of $84,000$ separate price forecasts
are made.}
Forecasts were obtained from each of the 
following methods:
\begin{itemize}
\item[(i)] {\em Na\"ive 1:} The price at the same time on the last 
observed day
is used as a benchmark point forecast.
\item[(ii)] {\em Na\"ive 2:} The average price 
at the same time of the day 
over the training sample
is used as a second benchmark point forecast.
\item[(iii)] {\em Fundamental:} The ensemble
of the marginal expectations from our regression models,
$\bm{\mu}^{\mbox{\tiny Ens}}_{T+h}=
\sum_{i=1}^r W_i \bm{\mu}^{(i)}_{T+h}$,
is used as a point forecast.
This is computed assuming the values of supply and flows
in the forecast period are known, so that $\bm{\mu}^{(i)}_{T+h}=
\bm{\eta}_{i,T+h}+E(\bm{\epsilon}_{i,T+h})$, where the marginal expectation 
$E(\bm{\epsilon}_{i,T+h})=(\tilde \alpha_{i,1},\ldots,\tilde \alpha_{i,r})'$ comprises
constants defined in Section~\ref{sec:smrp}. This approach treats the five price series as both serially and cross-sectionally
independent, conditional upon supply and flow observations.
\item[(iv)] {\em Copula \& Fundamental 1:} The ensemble distribution at Equation~(\ref{eq:ens1}),
with its mean $\bm{\pi}_{T+h|T}^{\mbox{\tiny Ens}}$ used as a 
point forecast.
This forecast distribution is computed assuming the values of supply and flows
in the forecast period are known. The approach differs from that above 
in that it exploits any serial and cross-sectional dependence captured by the copula model. 
\item[(v)] {\em Copula \& Fundamental 2:} The same ensemble distribution and point
forecast as above, 
but with supply and flows
in the forecast period determined
via hour-by-hour optimizations as outlined in Section~\ref{sec:jntfl}.
\item[(vi)] {\em Copula:} The copula multivariate time series model, but
without the regressions outlined in Equation~(\ref{eq:stat1}). 
Instead, the marginal distribution of each price series
is modeled only by its empirical distribution
function, so that no structural information is exploited.
\end{itemize}

To summarize the overall level
of accuracy,
we consider forecasts of the
demand-weighted (log) price across all five regions,
\[
\pi^{\mbox{\tiny DW}}_t=\sum_{l=1}^5 \left( \frac{d_{l,t}}{\sum_{j=1}^r d_{j,t}} \right) 
\pi_{l,t}\,,
\]
forecasts of which are constructed from the five regional forecasts
$\pi_{1,t},\ldots,\pi_{5,t}$.
The point forecast accuracy is measured using the mean
absolute forecast error (MAFE). The
means are computed over all forecasts, broken down by
different ranges of the forecast horizon, and reported
in Table~\ref{tab:mafe}. For example, the MAFE at horizons
of 4, 5 and 6 hours ahead is 0.983 for the `Copula' method, and is
an average of
$(3\times 100)$ AFE values.\footnote{The corresponding measures of 
accuracy of the individual regional price forecasts, and also for very high prices (which we define to be prices above the 95th percentile), are reported
in Tables 1--6 of the Supplementary Material.}

A number of insights can be drawn from the results. 
First, the two na\"ive approaches are dominated in most circumstances
by the statistical methods, highlighting the value of a model-based approach.
Second,
the extension of the `Fundamental' method to include multivariate serial dependence
via the copula model (ie. the `Copula \& Fundamental~1' method),
improves the accuracy
up to a horizon of 2 days ahead.
Third, for horizons of 3 or more days, 
the forecasts from the `Fundamental' method dominate all alternatives, highlighting 
the potential value of incorporating the structural information in
forecasts at a long horizon.
Fourth, forecasting interconnector flows (and supply) using optimization (ie. 
the `Copula \& Fundamental~2' method), reduces price forecast accuracy. Nevertheless, the
forecasts 
dominate the two na\"ive approaches for short horizons of 1 and 2 hours. 
The limited accuracy of this approach suggests that it would be worthwhile considering alternative approaches to forecasting interconnector flows. Fifth, 
the multivariate copula model without structural information (ie. the `Copula' method) 
performs particularly well for horizons of up to 2 days, but is dominated over longer 
horizons by the 
`Fundamental' method that incorporates such structural information.
The high accuracy of the copula multivariate time series model at short horizons
here mirrors that documented by Smith \& Vahey~(2016) for macroeconomic variables. Last, we note that only point forecast accuracy is considered here. Given the 
asymmetry in the logarithm of prices, it is worthwhile to extend the study 
to consider density forecasts from the statistical models and their accuracy.

\section{Discussion}\label{sec:discuss}
In this paper we use a spatial equilibrium model of price formation
to motivate multivariate
statistical models for a vector of regional electricity prices. There is a separate
multivariate model for each supply region, and the estimates and predictions
from each model can be combined into an ensemble. 
Key features of each multivariate statistical model include marginal
regressions with monotonic functions, and a high-dimensional
Gaussian copula to capture additional serial and cross-sectional dependence.
In estimating each regression model we employ a 
finite mixture of Gaussians for the disturbances, which is motivated by
the three price equilibria in the economic model. However, a nonparametric
disturbance using an infinite mixture model of the type popular in
the Bayesian literature (Hjort et al.~2010) 
could also be used.
When estimating the copula models, we use a two stage estimator. First, the
marginal distributions of the mean-corrected prices are estimated
using their empirical distribution functions. Second, the copula parameters
are estimated using maximum likelihood. Joint estimation of a Gaussian
copula and nonparametric marginals is far from 
straightforward (e.g. see Rosen \& Thompson~2015), 
and would be even more difficult joint
with the monotonic functions. For this reason two-stage estimators
are the most common approach for copula models with complex
margins (Joe~2005). 

Our econometric model of electricity prices 
has two main novel features. First, it is a multivariate model
that incorporates structural relationships
of inter-regional price formation. A number of recent 
statistical models of electricity prices employ
fundamental variables;
for example, see
Karakatsani \& Bunn~(2008a) and Gonz\'alez et al.~(2012). 
However, our model is 
the first study of which we are aware where the fundamental 
variables and their functional forms are
motivated by 
a network economic model.
The second novel aspect is that we employ
a copula model
for additional dependence in electricity prices. 
Smith, Gan \& Kohn~(2012) and Ignatieva \& Tr\"uck~(2016) recently employ a 5-dimensional copula to capture 
the dependence between prices in the NEM, while Manner, T\"urk \& Eichler~(2016) employ
a 4-dimensional copula to capture the dependence between price spike incidence in four regions in the NEM.
However, these low-dimensional copulas capture only cross-sectional dependence,
whereas we employ a parsimonious $5T$-dimensional copula
to capture both cross-sectional and serial dependence, which 
offers substantial advantages here. These include the ability 
to model the marginal distributions flexibly, so that the forecast densities
are substantially more realistic than those produced from overly simplistic
parametric distributions, such as the log-normal. While quantile
regression or time series models (e.g. Bunn et al.~2016)
also produce similarly flexible forecast densities,
they are harder to extend to multiple price series than the copula model.
And while we focus on the Gaussian copula here, it is possible to further 
extend
our analysis to vine copulas as in Smith~(2015). 

Electricity prices in the Australian market are highly dependent across 
regions, but
do not follow the law of one price. This is in part due to the unique
physical aspects of energy generation, such as interconnector constraints and
differing ramping times for generators; for example, see 
Clements~(2017) for a discussion of the effect of 
interconnector constraints on prices in the NEM, and Higgs, Lien \& Worthington~(2015)
for a discussion on the role of production capacity and generation mix.
However, prices are also likely to be affected by strategic bidding
and the exercise of market power by utilities; for example, see
theoretical work by Bunn \& Gianfreda~(2010) and a
recent empirical analysis by Apergis, Barun\'{i}k \& Lau~(2017).
Our empirical findings are consistent with these observations, identifying
both strong cross-sectional
and serial dependence in prices over-and-above that induced by
supply-side relationships in the marginal regressions.
We illustrate the usefulness of our
model in 
two event studies. The first is a 
supply-side shock equivalent to the outage of a major generator,
while
the second is a
price impulse in one region. We find
that such shocks have an impact on prices in all regions in
the NEM over a horizon of up to one week. This highlights
the importance
of employing a multivariate model in studies of regional prices in wholesale
markets with nodal pricing. Our validation study illustrates that our
model also shows promise when used to forecast prices. In comparison
to simple benchmarks,
the fundamental model 
that accounts for supply-side factors provides improved forecasts 
at longer horizons, while the inclusion of the time series copula model
increases accuracy at horizons under two days.

Finally, we note two avenues for future research. First,
the extent of inter-regional 
dependence between the variance, skewness and kurtosis of 
prices, along with the incidence of extreme prices (ie. price
spikes), has attracted much current
interest; for example, see
Lindstr\"om \& Regland~(2012), Aderounmu \& Wolff~(2014),
 Manner, Tr\"uk \& Eichler~(2016),
Apergis, Barun\'{i}k \& Lau~(2017) and
Han, Kordzakhia \& Tr\"uck~(2017).
Smith \& Vahey~(2016) show that the
Gaussian copula model employed here
can produce predictive
distributions with time-varying variance, skewness, kurtosis and
tail probabilities. The 
extent to which our proposed model captures inter-regional 
dependence in these moments and tail probabilities
is an interesting question. Second, while we document the accuracy of the point forecasts in a validation study, we do not document the density forecasts from
our statistical models. Given the asymmetry in the (log) prices, measuring the accuracy of these is also of great practical interest.
\newpage
\appendix

\noindent
{\bf \Large{Appendix}}

\noindent
This appendix provides further details on the monotonic smoothing method
in the presence of a mixture of normals outlined in Section~\ref{sec:mfe}.
Dropping the regression subscripts $i$ and $j$ for notational convenience, let
 $\bm{\omega}=(\omega_1,\omega_2,\omega_3)$,
 $\bm{\alpha}=(\alpha_1,\alpha_2,\alpha_3)$,
 $\bm{\sigma}^2=(\sigma_1^2,\sigma_2^2,\sigma_3^2)$, 
and $\bm{M}=(M_1, \ldots, M_T)$. Also, to keep the notation manageable, we will assume there are no cost functions $c_a$ in the model. The cost functions are estimated similarly to (and joint with) the supply function $S$.

To estimate $S$, we follow Shively et al. (2009) and define $X_J$ to 
consist of the columns of basis vectors $X=[\bm{b,b}^2, 
\ldots, (\bm{b}-\tilde b_m\bm{\iota})_+^2]$ that
correspond to the nonzero elements of $\bm J$. Therefore, the spline
in Equation~(\ref{eq:spline}) evaluated at the observed values is
$\mbox{\bf S}^{(m)}=X_J\bm{\beta}_J$. To make the model analytically tractable for use 
in an MCMC algorithm, we re-parameterize to give $\mbox{\bf S}^{(m)}=W_J\bm{\gamma}_J$, 
where $\bm{\gamma}_J=L_J\bm{\beta}_J$, $W_J=X_JL_J^{-1}$, and $L_J$ is the 
lower triangular matrix  defined in Section~\ref{sec:mfe}.
Note that the constrained prior for the $\bm{\beta}_J$ discussed in
Section~\ref{sec:mfe} induces a $N(0,c\sigma^2I_d)$ prior for $\bm{\gamma}_J$ 
constrained to the region $\bm{\gamma}_J>\bm{0}$,
where $d=\sum_{j=1}^{m+2}J_j$.

The posterior distribution for $\bm{M}, \bm{\omega}, \bm{\alpha}, 
\bm{\sigma}^2, \bm{J}, \bm{\gamma}|\bm{\pi}$ has density
$$f(\bm{M, \omega, \alpha, \sigma^2, J, \gamma|\pi}) \propto 
f(\bm{\pi|M, \alpha, \sigma}^2, \bm{J, \gamma})
f(\bm{M, \omega, \alpha, \sigma}^2, \bm{J, \gamma})$$
where
\begin{equation*}
f(\bm{\pi|M, \alpha, \sigma}^2, \bm{J, \gamma})=\prod_{l=1}^3\left\{ \prod_{t:M_t=l}(2\pi\sigma_l^2)^{-1/2}
\exp\left[ \frac{-1}{2\sigma_l^2}(\pi_t-\bm{w}_{Jt}
\bm{\gamma}_J-\alpha_l)^2  \right] \right\}, \label{eq2}
\end{equation*}
$\bm{w}_{Jt}$ is the t-th row of $W_J$, and 
$$f(\bm{M, \omega, \alpha, \sigma}^2, \bm{J, \gamma}) = 
\mbox{Pr}(\bm{M|\omega})f(\bm{\omega})f(\bm{\alpha})f(\bm{\sigma}^2)f(\bm{\gamma|J})\mbox{Pr}(\bm{J})$$ 
is the prior distribution defined in Section~\ref{sec:mfe}. Letting
\begin{equation*}
s(\bm{\pi,M, \alpha, \sigma}^2, \bm{J, \gamma})=-\log[f(\bm{\pi|M, \alpha, \sigma^2, J, \gamma})],
\end{equation*}
and following Shively et al.~(2011), we introduce a
scalar latent variable $z$ such that
\begin{equation}\tag{A1}
f(\bm{M, \omega, \alpha, \sigma}^2, \bm{J, \gamma}, z|\bm \pi) \propto e^{-z} I\left(z > s(\bm{\pi,M, \alpha, \sigma}^2, \bm{J, \gamma}) \right)f(\bm{M, \omega, \alpha, \sigma}^2, \bm{J, \gamma})\,.  \label{eq:append1}
\end{equation}

The sampling scheme below is used to generate Monte Carlo iterates from the posterior
distribution, augmented with this latent variable. For notational purposes,
let $\bm{\omega}_{(-l)}$, $\bm \alpha_{(-l)}$, $\bm \sigma_{(-l)}^2$, $\bm J_{(-l)}$, and  $\bm \gamma_{(-l)}$ 
represent $\bm \omega$, $\bm \alpha$, $\bm \sigma^2$, $\bm J$ and $\bm \gamma$,
respectively, without the $l$-th element.

\vspace{0.1in}
\indent
Step 0: Start with some initial values of $\bm M^{[0]},\bm \omega^{[0]},\bm \alpha^{[0]}, (\bm \sigma^2)^{[0]},\bm J^{[0]}$, $\bm \gamma^{[0]}$;

\indent
Step 1: Generate $z$ conditional on $\bm{\pi, M, \alpha, \sigma}^2, \bm{J}$, $\bm \gamma$;

\indent
Step 2: For $t=1,\ldots,T$, generate $M_t$ conditional on $\bm \pi, z, \bm M_{(-t)}$, $\bm \omega$, $\bm \alpha$, $\bm \sigma^2$, $\bm J$, $\bm \gamma$;

\indent
Step 3: Generate $\bm \omega$ conditional on $\bm{M}$;

\indent
Step 4: For $l=1,2,3$, generate $\alpha_l$ conditional on $\bm{\pi}, z, \bm{M}$, $\bm \alpha_{(-l)}$, $\bm \sigma^2$, $\bm J$, $\bm \gamma$;

\indent
Step 5: For $l=1,2,3$, generate $\sigma^2_l$ conditional on $\bm \pi$, $z$, $\bm M$, $\bm \alpha$, $\bm \sigma^2_{(-l)}$, $\bm J$, $\bm \gamma$; and

\indent
Step 6: For $j=1, \ldots, m+2$, generate $(J_j,\gamma_j)$ jointly conditional $\bm \pi$, $z$, $\bm{M}$, $\bm \alpha$, $\bm \sigma^2$, $\bm J_{(-j)}$, $\bm \gamma_{(-j)}$.

\vspace{0.1in}

At Step~1,
generate $z^* \sim$ Exp(1) and compute $z=z^* + s(\bm\pi,\bm{M},\bm \alpha,\bm \sigma^2,\bm J,\bm \gamma)$. At Step~2,
generate $M_t$ from a multinomial distribution with
\[
\mbox{Pr}(M_t=l|\bm \pi, z, \bm{M}_{(-t)}, \bm \omega, \bm \alpha, \bm \sigma^2, \bm J,\bm \gamma)
\propto \left\{\begin{array}{ccl} \omega_l &\mbox{if} &s(\bm \pi,M_t,\bm{M}_{(-t)},\bm \alpha,\bm \sigma^2,\bm J,\bm \gamma)-z<0 \\ 
 0 &\mbox{if} &s(\bm \pi,M_t,\bm{M}_{(-t)},\bm \alpha,\bm \sigma^2,\bm J,\bm \gamma)-z \ge 0\,. \end{array}\right. 
\]
At Step~3, 
generate
$\bm \omega \sim$ Dirichlet$(n_1+1,n_2+1,n_3+1)$, where $n_l$ is the number of observations with $M_t=l$ (ie. from component $l$).

At Step~4,
to keep the notation manageable, consider generating $\alpha_1$.  
Let $b_L$ and $b_U$ represent the roots with respect to $\alpha_1$ of the function
$s(\bm \pi,\bm{M},\alpha_1, \bm \alpha_{(-1)},\bm \sigma^2,\bm J,\bm \gamma)-z$ 
in the indicator function in Equation~(\ref{eq:append1}). 
Then $\alpha_1|\bm\pi, z, \bm{M}, \bm \alpha_{(-1)}, \bm \sigma^2, \bm J, \bm \gamma \sim N(0,c)$,
constrained to the interval $\max(\alpha_2,b_L)<\alpha_1<\min(\alpha_3,b_U)$.
The values
$\alpha_2$ and $\alpha_3$ are generated similarly. 

At Step~5, 
consider generating $\sigma^2_1$. 
Using a technique similar to Step~4, let $b_L$ and $b_U$ represent the roots of
$s(\bm \pi,\bm{M},\bm \alpha,\sigma_1^2,\bm \sigma_{(-1)}^2,\bm J,\bm \gamma)-z$ 
with respect to $\sigma^2_1$. Then $\sigma^2_1|\bm \pi, z, \bm{M}, \bm \alpha, \bm \sigma^2_{(-1)}, \bm J, \bm \gamma \sim \mbox{Uniform}(a_L,a_U)$,
where $a_L = \max(0,b_L)$ and $a_U = \min(0.25\sigma^2_{(2)},b_U)$.

Following Shively et al. (2010), at Step~(6) $(J_j,\gamma_j)$ is generated as a pair by generating $J_j$ first, and then $\gamma_j|J_j$. 
To begin, note that
$$f(J_j,\gamma_j|\mbox{---},z) \propto I[s(\mbox{---},J_j=1,\gamma_j)-z<0]
f(\gamma_j|J_j)\mbox{Pr}(J_j)$$
where `$\mbox{---}$' denotes 
$(\bm \pi, \bm M,\bm \alpha,\bm \sigma^2, \bm J_{(-j)},\bm \gamma_{(-j)})$.
If $s(\mbox{---},J_j=0)-z>0$,
then $\mbox{Pr}(J_j=0|\mbox{---},z)=0$. Otherwise, $\mbox{Pr}(J_j=0|\mbox{---},z) \propto \mbox{Pr}(J_j=0)$.

To find $\mbox{Pr}(J_j=1|\mbox{---},z)$, note that if $s(\mbox{---},J_j=1,\gamma_j)-z>0$ 
for all $\gamma_j>0$, 
then $\mbox{Pr}(J_j=1|\mbox{---},z)=0$. Otherwise, let $R$ represent the region of $\gamma_j$ values where 
$s(\mbox{---},J_j=1,\gamma_j)-z<0$. Then
$$f(J_j=1,\gamma_j|\mbox{---},z) \propto I(\gamma_j \in R)f(\gamma_j|J_j=1)\mbox{Pr}(J_j=1)$$
and 
$$\mbox{Pr}(J_j=1|\mbox{---},z) \propto \left[\int_Rf(\gamma_j|J_j=1)\mbox{d}\gamma_j\right]\mbox{Pr}(J_j=1).$$
If $J_j=1$, then $\gamma_j$ is generated from a $N(0,c\sigma^2)$ distribution constrained to the region $R$.

Let $\{\bm J^{[r]},\bm \gamma^{[r]}\}_{r=1}^R$ be the iterates of $(\bm J, \bm \gamma)$ in the sampling period. 
Then an estimate of the posterior mean of $S^{(m)}(b_{t})$, and therefore an estimate of $S(b_t)$, is 
$\frac{1}{R}\sum_{r=1}^R \left[ \bm w_{J^{[r]}t}\bm \gamma^{[r]}_{J^{[r]}} \right]$.
The estimates of the posterior means of $\omega_l$, $\alpha_l$, $\sigma^2_l$ and $\mbox{Pr}(M_t=l|\bm \pi)$ are obtained similarly.
%

\newpage
\onehalfspacing
\section*{References}
\parindent=0pt              
\begin{trivlist}

\item[]
Aderounmu, A. A. \& Wolff, R., (2014). 
`Modeling dependence of price spikes in Australian electricity markets',
{\em Energy Risk}, 11, 60-–65. 
\item[] 
Apergis, N., J. Barun\'{i}k and M.C.K. Lau, (2017).
`Good volatility, bad volatility: What drives the asymmetric connectedness of Australian electricity markets?', {\em Energy Economics}, 66, 108--115.
\item[]
Australian Electricity Management Organisation, (2010). 
{\em An Introduction to Australia's Electricity Market}, 
accessed at \underline{www.aemo.gov.au}. 
\item[]
Australian Energy Regulator, (2011). {\em State of the Energy Market 2011},
accessed at \underline{www.accc.gov.au}. 
\item[]
Bakirtzis, A.G., (2001). `Aumann-Shapley transmission congestion pricing',
{\em Power Engineering Reivew, IEEE}, 21, 67--69.
\item[]
Bedford, T. and R.M. Cooke, (2001). `Probability density decomposition for conditionally 
dependent random variables modeled by vines', {\em Annals of Mathematics and Artificial Intelligence},
32, 245--268.  
\item[]
Biller, B. (2009). `Copula-Based Multivariate Input Models for Stochastic Simulation',
{\em Operations Research}, 57, 4, 878--892.
\item[]
Bunn, D.W. \& A. Gianfreda, (2010). `Integration and shock transmissions
across European electricity forward markets', {\em Energy Economics},
32, 278--291.
\item[]
Bunn, D.W., A. Andresen, D. Chen \& S. Westgaard, (2016).
`Analysis and forecasting of electricity price risks with quantile
factor models', {\em The Energy Journal}, 37(1).
\item[]
Clements, A.E., A.S. Hurn \& Z. Li. (2016).
`Forecasting day-ahead electricity load using a multiple equation
time series approach.'
{\em European Journal of Operational Research}, 251, 522--530.
\item[]
Clements, A.E., A.S. Hurn, \& Z. Li. (2017).
`The Effect of Transmission Constraints on Electricity Prices',
{\em The Energy Journal}, 38(4).	
\item[]
De Menezes, L.M., D.W. Bunn \& J.W. Taylor~(2000). 
`Review of Guidelines for the Use of Combined Forecasts',
{\em European Journal of Operational Research}, 120, 190--204.
\item[]
DeVany, A.S. \& W.D. Walls, (1999). `Cointegration analysis of spot electricity
prices: insights on transmission efficiency in the western US', 
{\em Energy Economics}, 21, 435--448.
\item[]
Ferkingstad, E., A. L{\o}land \& M. Wilhelmsen~(2011). `Causal modeling and
inference for electricity markets', {\em Energy Economics}, 33, 404--412.
\item[]
Fr\"uhwirth-Schnatter, S., (2006). {\em Finite Mixture and Markov Switching
Models}, Springer: NY.
\item[]
Geman, H. \& A. Roncoroni, (2006). `Understanding the Fine Stucture of 
Electricity Prices', {\em The Journal of Business}, 79: 3, 1225--1261.
\item[]
Gonz\'alez, V., J. Contreras \& D. Bunn, (2012). `Forecasting
Power Prices using a Hybrid Fundamental-Econometric Model',
{\em IEEE Transations on Power Systems}, 27, 363--372.
\item[]
Haldrup, N. \& M. \O. Nielsen, (2006).
`A regime switching long memory model for electricity prices',
{\em Journal of Econometrics}, 135, 349--376.
\item[]
Han, L., N. Kordzakhia \& S. Tr\"uck, (2017). `Volatility Spillovers in Australian Electricity Markets', Working Paper. 
\item[]
Harvey, A. \& S.J. Koopman, (1993). 
`Forecasting Hourly Electricity Demand Using Time-Varying Splines',
{\em Journal of the American Statistical Association}, 88, 1228--1236.
\item[]
Higgs, H., G. Lien \& A.C. Worthington, (2015). `Australian evidence on the role of interregional flows, production capacity, and generation mix in wholesale electricity prices and price volatility', {\em Economic Analysis and Policy}, 48, 172--181.
\item[]
Hjort, N.L., C. Holmes, P. M\"uller, S.G. Walker, (2010). {\em Bayesian 
Nonparametrics}, Cambridge Series in Statistical and Probabilistic Mathematics.
\item[]
Huisman, R. \& R. Mahieu, (2003). `Regime jumps in electricity prices',
{\em Energy Economics}, 25, 425--434.
\item[]
Ignatieva, K. \& S. Tr\"uck~(2016). `Modeling spot price dependence
in Australian electricity markets with applications to risk management',
{\em Computers and Operations Research}, 415--433.
\item[]
Janczura, J. \& R. Weron, (2010). `An emprical comparison of alternative
regime-switching models for electricity spot prices', 
{\em Energy Economics}, 32, 1059--1073. 
\item[]
Joe, H. (2005). 
`Asymptotic efficiency of the two-stage estimation method for copula-based
models', {\em Journal of Multivariate Analysis}, 94, 401--419.
\item[]
J\'onsson, T., P. Pinson, H. Madsen \& H. Nielsen,~(2014).
`Predictive Densities for Day-Ahead Electricity Prices Using Time-Adaptive
Quantile Regression', {\em Energies}, 7, 5523--5547.
\item[]
Jore, A. S., J. Mitchell \& S. P. Vahey, (2010). `Combining
forecast densities from VARs with uncertain instabilities',
{\em Journal of Applied Econometrics}, 25, 621--634.
\item[]
Karakatsani, N. V. \& D.W. Bunn~(2008a). `Forecasting electricity prices:
The impact of fundamentals and time-varying coefficients', 
{\em International Journal of Forecasting}, 24, 764--785.
\item[]
Karakatsani, N. V. \& D.W. Bunn~(2008b). `Intra-day and regime-switching
dynamics in electricity price formation', {\em Energy Economics},
30, 1776-1797.
\item[]
Kirschen, D.S., (2003). `Demand-Side View of Electricity Markets', {\em IEEE
Transactions on Power Systems}, 18: 2, 520--527.
\item[]
Kirschen, D.S. \& G. Strbac, (2004). {\em Fundamentals of Power Systems 
Economics}, Wiley.
\item[]
Knittel, C.R., \& M.R. Roberts, (2005). `An empirical examination of 
restructured electricity prices', {\em Energy Economics}, 27, 791--817.
\item[]
Koop, G., M.H. Pesaran \& S.M. Potter, (1996). `Impulse response analysis
in nonlinear multivariate models', {\em Journal of Econometrics}, 74, 119--147.
\item[] 
Lindstr\"om, E. and F. Regland, (2012), `Modeling extreme dependence between European electricity markets', {\em Energy Economics}, 34, 899--904.
\item[]
Liu, S.D., J.B. Jian, and Y.Y. Wang, (2010). 
`Optimal dynamic hedging of electricity futures based on copula-GARCH models',
in {\em Industrial Engineering and Engineering Management (IEEM), 
2010 IEEE International Conference on}, 2498--2502.
\item[]
L\"{u}tkepohl, H., (2006). {\em New Introduction to Multiple Time Series
Analysis}, Springer.
\item[]
Manner, H., D. T\"urk and M. Eichler, (2016). 
`Modeling and forecasting multivariate electricity price spikes'
{\em Energy Economics}, 60, 255--265.
\item[]
Mauricio, J.A. (1995). `Exact Maximum Likelihood Estimation of Stationary
Vector ARMA Models', {\em Journal of the American Statistical Association},
90, 282--291.
\item[]
Mitchell, J. \& Hall, S. (2005). `Evaluating, Comparing and Combining
Density Forecasts Using the KLIC with an Application to the Bank of
England and NIESR Fan Charts of Inflation',
{\em Oxford Bulletin of Economics and Statistics}, 67, 995--1033.
\item[]
Nagurney, A., (1999). {\em Network Economics: A Variational Inequality
Approach}, 2nd Ed.,
Kluwer Academic Publishers.
\item[]
Panagiotelis, A. \& M. Smith, (2008). `Bayesian density forecasting of intraday electricity
prices using 
using multivariate skew-t distributions',
{\em International Journal of Forecasting}, 24, 710--727.
\item[]
Papaefthymiou, G., \& Kurowicka, D. (2009). 
`Using copulas for modeling stochastic dependence in power system uncertainty
analysis', {\em IEEE Transactions on Power Systems}, 24, 40--49.
\item[]
Patton, A.J., (2006). `Modelling Asymmetric Exchange Rate Dependence',
{\em International Economic Review}, 47, 527--556.
\item[]
Ramanathan, R., R. Engle, C.W. Granger, F. Vahid-Araghi,
\& C. Brace, (1997). `Short-run forecasts of electricity
loads and peaks', {\em International Journal of Forecasting}, 13,
161--174.
\item[]
Rosen, O. \& W.K. Thompson, (2015). `Bayesian semiparametric copula estimation 
with application to psychiatric genetics', {\em Biometrical Journal}, 57, 
468--484.
\item[]
Rosenblatt, M. (1952). `Remarks on a multivariate transform', {\em Annals of 
Mathematical Statistics}, 23, 470--472.
\item[]
Shih, J.H. \& T.A. Louis, (1995). `Inference on the Association Parameter
in Copula Models for Bivariate Survival Data', {\em Biometrics}, 51, 1384--1399.
\item[] 
Shimazaki, H. \& S. Shinomoto, (2010). `Kernel bandwidth optimization in spike rate estimation', {\em Journal of Computational Neuroscience}, 29, 171--182.
\item[]
Shively, T.S., T.W. Sager \& S.G. Walker, (2009).
`A Bayesian approach to non-parametric monotone function estimation',
{\em Journal of the Royal Statistical Society}, Series B, 71, 159--175.
\item[]
Shively, T.S., S.G. Walker \& P. Damien, (2011).
`Nonparametric function estimation subject to monotonicity, convexity and 
other shape constraints', {\em Journal of Econometrics}, 161, 166--181.
\item[]
Sloughter, J.M., T. Gneiting, \& A. Raftery,~(2010).
`Probablistic Wind Speed Forecasting Using Ensembles and Bayesian
Model Averaging', {\em Journal of the American Statistical Association},
105, 25--35.
\item[]
Smith, M. (2000). `Modeling and short-term forecasting of New South Wales electricity system load',
{\em Journal of Business \& Economic Statistics}, 18, 465--478.
\item[]
Smith, M.S., (2015). `Copula Modeling of Dependence in Multivariate 
Time Series', {\em International Journal of Forecasting}, 31, 815--833.
\item[]
Smith, M.S., Q. Gan, \& R. Kohn, (2012). `Modeling dependence using 
skew t copulas: Bayesian inference and applications', 
{\em Journal of Applied Econometrics}, 27, 500--522. 
\item[]
Smith, M. \& R. Kohn, (1996). `Nonparametric regression via Bayesian
variable selection', {\em Journal of Econometrics}, 75, 317--343.
\item[]
Smith, M., A. Min, C. Almeida, \& C. Czado.~(2010). 
`Modeling longitudinal data using a pair-copula decomposition of 
serial dependence', {\em Journal of the American Statistical Association},
105, 492, 1467--1479.
\item[]
Smith, M.S. \& S. Vahey, (2016). `Asymmetric density forecasts for 
U.S. macroeconomic variables from a Gaussian copula model of 
cross-sectional and serial dependence', 
{\em Journal of Business and Economic Statistics}, 34, 416--434. 
\item[]
Song, P., (2000).
`Multivariate Dispersion Models Generated from Gaussian
Copula', {\em Scandinavian Journal of Statistics}, 27, 305--320.
\item[]
Taylor, J.W., L. de Menezes \& P. McSharry,~(2006).
`A comparison of univariate methods for forecasting electricity
demand up to a day ahead', {\em International Journal of Forecasting},
22, 1--16.
\item[]
Wang, X., J. Cai \& K. He.~(2015).
`EMD Copula based Value at Risk Estimates for Electricity Markets'
{\em Procedia Computer Science}, 55, 1318--1324.
\item[]
Weron, R., M. Bierbrauer \& S. Tr\"uck~(2004). 
`Modeling electricity prices: jump diffusion and regime switching'
{\em Physica A: Statistical Mechanics and its Applications},
336, 39--48.
\item[]
Weron, R. \& Misiorek, A. (2008). 
`Forecasting spot electricity prices: A comparison of parametric and semiparametric time series models',
{\em International Journal of Forecasting}, 24, 744-763.
\item[]
Weron, R. (2014). `Electricity price forecasting: A review of the state-of-the-art
with a look to the future', {\em International Journal of Forecasting},
30: 1030--1081.
\item[]
Westgaard S. (2014). `Energy Spread Modeling',
in Prokopczuk, M. {\em Energy Pricing Models: Recent Advances, Methods, and Tools}, Palgrave Macmillan.	
\item[]
Wolak, F., (2007). `Quatifying the supply-side benefits from
forward contracting in wholesale electricity markets', 
{\em Journal of Applied Econometrics}, 22, 1179--1209.
\item[]
Wood, A.J. \& B.F. Wollenberg, (1995). {\em
Power Generation, Operation and Control},
2nd Ed., N.Y.: Wiley.
\item[] 
Ziel, F. \& Steinert, R. (2016). `Electricity price forecasting using sale and purchase curves: The X-Model', 
{\em Energy Economics}, 59, 435--454.
\end{trivlist}

\pagestyle{empty}
\newpage
\begin{table}[htbp]
\begin{center}
\begin{tabular}{lrrrrr} \hline \hline
Region &Coal &Gas &Hydro &Other &Total \\ \hline
NSW &11.8	&2.2	&2.7	&0.6	&17.3\\
QLD &8.5	&3.1	&0.6	&0.8	&13.1\\
VIC &6.6	&1.9	&2.2	&0.5	&11.2\\
SA  &0.7	&2.2	&0.0	&1.8	&4.8\\
TAS &0.0	&0.6	&2.3	&0.2	&3.0\\ \hline\hline
\end{tabular}
\end{center}
\caption{Registered generation capacity (GW) 
by regions and fuel source during 2011. Source: AER, {\em State of the Energy 
Market}, p.28.}
\label{tab:capacity}
\end{table}

\begin{table}[htbp]
	\begin{center}
		\begin{tabular}{lcccccccccccc} \hline \hline
			Arc &$v_1$ &$v_2$ &$v_3$ &$v_4$ &$v_5$ &$v_6$ \\ 
			Interconnector &Terr &Terr &QNI &QNI &V-N &V-N \\ 
			Origin &QLD &NSW &QLD &NSW &NSW &VIC\\
			Destination &NSW &QLD &NSW &QLD &VIC &NSW\\ 
			Nominal &180/230$^\dagger$ &180/230$^\dagger$ &1078& 700& 1,350 &1,550 \\
			Mean &104 &0 &736 &3 &87 &475\\
			Max. &231 &50 &1078 &410 &1,348 &1,525\\ \hline
			Arc &$v_7$ &$v_8$ &$v_9$ &$v_{10}$ &$v_{11}$ &$v_{12}$ \\
			Interconnector &Hey &Hey &Mry &Mry &Bass &Bass\\ 
			Origin &SA &VIC &SA &VIC &VIC &TAS \\
			Destination &VIC &SA &VIC &SA &TAS &VIC \\
			Nominal &460 &460 &220 &220 &480 &600 \\
			Mean &35 &133 &31 &10 &140 &132\\
			Max. &455 &457 &172 &220 &478 &594\\ \hline\hline
		\end{tabular}
	\end{center}
	\caption{Summaries of half-hourly directed transmission flows (MWh)
		for the interconnectors
		in the NEM depicted in Figure~1. Each column corresponds to a different flow.
		Note that the minimum flows are zero, and that the mean and maximum flows
		exhibit high asymmetry with regards to direction. Nominal upper capacity
		is also given, although actual capacity constraints differ at any given 
		point in time depending upon a number of factors.\newline $^\dagger$The Terranora 
		interconnector can operate in an overload mode of 230MWh for
		short periods of time.}
	\label{tab:va}
\end{table}

\begin{table}[htbp]
	\begin{center}
		\begin{tabular}{lccccc} \hline \hline
			&NSW &QLD &SA &TAS &VIC \\ \hline
			&\multicolumn{5}{c}{{\em Half-hourly Prices (\$/MWh)}} \\ 
			Mean Peak &53.77  &42.44  &60.16 &35.79 &45.28 \\
			Min. Peak &-264.31 &-506.75 &-658.68 &-409.48 &-563.03 \\
			Max. Peak &12,136 &9,044  &12,200 &12,400 &9,999 \\
			Mean Off-Peak &24.40  &19.18  &21.55  &26.70  &21.41\\
			Min. Off-Peak &0 &-1,000 &-997 &-464 &-817 \\ 
			Max. Off-Peak &6,267  &302 &119 &12,400 &118\\
			No. $>$500 &46 &38 &37 &27 &32 \\
			No. $<$0 &6 &39 &149 &65 &14 \\ \hline
			&\multicolumn{5}{c}{{\em Half-hourly Supply (GWh) }} \\ 
			Mean Peak  &8.63 &7.30 &1.61 &1.32 &6.70 \\
			Min. Peak  &5.06 &5.81 &0.78 &0.47 &4.51\\
			Max. Peak  &13.01 &9.12 &2.97 &2.19 &9.36\\
			Mean Off-Peak &6.72 &6.30 &1.37 &0.99 &6.00 \\
			Min. Off-Peak &3.82 &4.58 &0.66 &0.40 &4.36\\
			Max Off-Peak  &12.04 &8.53 &2.64 &2.15 &8.19\\ \hline \hline
		\end{tabular}
	\end{center}
	\caption{Summaries of half-hourly price data (in Australian dollars per megawatt hour) and 
		supply data (in gigawatt hours = 1000 megawatt hours) for the five regions
		of the NEM. The summaries are computed for periods of
		both peak demand (09:00--20:00 Sydney time)
		and off-peak demand (20:30--08:30 Sydney time). The number of times
		prices exceed 500 \$/MWh, or are below 0 \$/MWh, are also reported.} 
	\label{tab:pib}
\end{table}

\begin{table}[htbp]
	\begin{center}
		\begin{tabular}{lccccc} \hline \hline
			&NSW &QLD &SA & TAS &VIC \\ \hline
			&\multicolumn{5}{c}{Component 1}\\
			$\alpha_1$ &6.928 &6.924 &6.926 &6.929 &6.928 \\
			&{\footnotesize (6.927,6.930)} &{\footnotesize (6.923,6.924)} &{\footnotesize (6.926,6.926)} &{\footnotesize (6.928,6.929)} &{\footnotesize (6.928,6.928)}\\
			$\sigma_1$ &0.0025 &0.0035 &0.0056 &0.0053 &0.0059 \\
			&{\footnotesize (0.0024,0.0026)} &{\footnotesize (0.0035,0.0036)} &{\footnotesize (0.0055,0.0056)} &{\footnotesize (0.0052,0.0053)} &{\footnotesize (0.0058,0.0059)}\\
			$\omega_1$ &0.751  &0.864 &0.901 &0.900 &0.933\\
			&{\footnotesize (0.735,0.771)} &{\footnotesize (0.859,0.870)} &{\footnotesize (0.894,0.907)} &{\footnotesize (0.894,0.906)} &{\footnotesize (0.929,0.937)}\\ \hline
			&\multicolumn{5}{c}{Component 2}\\
			$\alpha_2$ 
			&6.928  &6.924  &6.876 &6.928 &6.928\\
			&{\footnotesize (6.927,6.930)} &{\footnotesize (6.923,6.924)} &{\footnotesize (6.809,6.921)} &{\footnotesize (6.928,6.929)} &{\footnotesize (6.928,6.928)}\\
			$\sigma_2$
			&0.012 &0.023 &1.251 &0.024 &0.033\\
			&{\footnotesize (0.011,0.013)} &{\footnotesize (0.022,0.024)} &{\footnotesize (1.094,1.417)} &{\footnotesize (0.023,0.025)} &{\footnotesize (0.032,0.035)}\\
			$\omega_2$
			&0.222 &0.123 &0.008 &0.094 &0.063\\
			&{\footnotesize (0.206,0.236)} &{\footnotesize (0.118,0.128)} &{\footnotesize (0.007,0.009)} &{\footnotesize (0.088,0.100)} &{\footnotesize (0.059,0.067)}\\ \hline
			&\multicolumn{5}{c}{Component 3}\\
			$\alpha_3$
			&7.092 &7.160 &6.941 &7.343 &7.646 \\
			&{\footnotesize (7.038,7.171)} &{\footnotesize (7.097,7.224)} &{\footnotesize (6.939,6.942)} &{\footnotesize (7.256,7.430)} &{\footnotesize (7.495,7.800)} \\
			$\sigma_3$ 
			&0.446 &0.732 &0.024 &0.670 &0.909\\
			&{\footnotesize (0.416,0.486)} &{\footnotesize (0.686,0.781)} &{\footnotesize (0.023,0.025)} &{\footnotesize (0.609,0.734)} &{\footnotesize (0.805,1.021)}\\
			$\omega_3$ 
			&0.027 &0.013 &0.091 &0.006 &0.004\\
			&{\footnotesize (0.022,0.031)} &{\footnotesize (0.011,0.014)} &{\footnotesize (0.086,0.097)} &{\footnotesize (0.005,0.007)} &{\footnotesize (0.003,0.004)}\\
			\hline\hline
		\end{tabular}
	\end{center}
	\caption{Posterior estimates 
		of the parameters of the three Gaussian mixture components from the
		five 
		VIC supply regressions. Results in each column correspond to the regressions
		with different regional prices. For each parameter the posterior mean is 
		given, along with the 90\% posterior probability interval in parentheses.} 
	\label{tab:margins}
\end{table}

\begin{sidewaystable}[htbp]
\begin{center}
\begin{tabular}{lcccccccccccc} \hline \hline
 &\multicolumn{12}{c}{Hours Ahead in Forecast Horizon} \\
Method or Model &1 &2 &3 &4--6 &7--12 &13--24 &25--48 &49--72 &73--96
&97--120 &121--144 &145--168 \\
 &\multicolumn{6}{c}{{\em ($\leq$ 1 day ahead)}} &{\em (2 days)} &{\em (3 days)} &{\em (4 days)}
&{\em (5 days)} &{\em (6 days)} &{\em (7 days)} \\
 \hline
Na\"{i}ve~1   &1.067  & 1.067  &1.066  &1.066  &1.068 &1.107  &1.503  &1.732  &1.855 &1.892 &1.962 &1.989\\ 
Na\"{i}ve~2   &1.304 &1.304 &1.304 &1.305& 1.314 &1.513 &1.604 &1.698 &1.769 &1.808 &1.809 &1.805\\ 
Fundamental &1.186 &1.186 &1.186 &1.186 &1.195 &1.379 &1.457 &1.538 &1.602 &1.639 &1.640 &1.636\\
 Copula & 0.737 &0.886 & 0.953 &0.983 &1.014 &1.157 &1.407 &1.553 &1.609 &1.639 &1.682 &1.753 \\ 
 Copula \& Fundamental 1 &0.792 &0.935 &1.005 &1.051 &1.084 &1.226 &1.456 &1.632 &1.726 &1.794  &1.897 &2.010\\ 
 Copula \& Fundamental 2 &0.848 &1.005 &1.074 &1.126 &1.169 &1.321 &1.556 &1.736 &1.835 &1.912 &2.020 &2.144\\ 
 \hline\hline
\end{tabular}
\end{center}
\caption{Summary of the accuracy of the point forecasts in the validation study. 
Each value corresponds to a mean absolute
forecast error (MAFE), multiplied by 100 to aid presentation. The means are computed over the 16,800
hourly forecasts, broken down by forecast horizon; the latter of which are reported in different columns. 
Each row corresponds to a forecasting method listed in Section~5.} 
\label{tab:mafe}
\end{sidewaystable}

\end{document}